\documentclass[aps,pra,twocolumn,groupedaddress, showpacs]{revtex4-1}
\usepackage{amsmath}
\usepackage{graphicx}
\makeatletter

\newcommand{\Rmnum}[1]{\expandafter\@slowromancap\romannumeral #1@}
\makeatother
\begin{document}

\title{Examining Resonant Inelastic Spontaneous Scattering of Classical Laguerre-Gauss Beams From Molecules}

\author{Aaron S. Rury}
\affiliation{Applied Physics Program, University of Michigan, Ann Arbor, MI 48109 USA\\
Jet Propulsion Laboratory, California Institute of Technology, Pasadena, CA, 91109, USA}
\email{arury@caltech.edu}
\date{\today}

\begin{abstract}This paper theoretically treats the spontaneous resonant inelastic scattering of Laguerre-Gauss (LG) beams from the totally symmetric vibrations of complex polyatomic molecules within the semi-classical framework. We develop an interaction Hamiltonian that accounts for the position of the molecule within the excitation beam to derive the effective differential scattering cross-section of a classical LG beam from a molecule using the frequency domain third order nonlinear optical response function. To gain physical insight into this scattering process, we utilize a model vibronic molecule to study the changes to this scattering process. For specific molecular parameters including vibrational frequency and relative displacement of the involved electronic states, this investigation shows that an incident LG beam asymmetrically enhances one of two participating excitation transitions causing modulation of the interference present in the scattering process. This modulation allows a pathway to coherent control of resonant inelastic scattering from complex, poly-atomic molecules. We discuss the possible application of this control to the resonant x-ray inelastic scattering (RIXS) of small poly-atomic molecules central to applications ranging from single molecule electronics to solar energy science. 
\end{abstract}


\maketitle

\section{Introduction}
\noindent As a first order approximation in solving the molecular Schrodinger equation, one treats the molecule's electronic and nuclear degrees of freedom on disparate time scales, allowing separation of their associated quantum mechanical wavefunctions as a tensor product.\cite{fischer1984vibronic} Such a treatment corresponds to an adiabatic approximation of the molecule's behavior, the most well-known of which is the Born-Oppenheimer approximation.\cite{BO} 

However, there are many cases in which a complete understanding of molecular behavior necessitates treating the coupling between the electronic and nuclear 'motions' of molecules. Interactions that call for relaxation of the adiabatic approximation are broadly known as vibronic coupling, indicating a coupling between the electronic and vibrational degrees of freedom of a molecule. Vibronic coupling is especially important in the context of light-molecule interactions and is necessary to understand the UV absorption spectra of aromatic molecules, the optical properties of transition metal complexes and even resonant x-ray interactions in molecules and molecular materials.\cite{harris1989symmetry,Skytt_PRL_1993,Zink_review} 

Despite the importance of vibronic coupling in determining molecular behavior and dynamics, little research has investigated the ability to control the coupling of electronic and vibrational molecular degrees of freedom. While many researchers have investigated how changes to the spectrum of an ultrashort pulse of light can control atomic and molecular dynamics, few researchers have looked into the ability of light fields, continuous or pulsed, to affect the fundamental coupling of nuclear and electronic degrees of freedom that determine subsequent dynamics and behavior.\cite{coherent_1,coherent_2,coherent_3,coherent_4,Weinacht_1,Weinacht_2,Weinacht_3} However, recent studies propose the possibility that tailored transverse modes of light beams control the coupling of internal degrees of freedom of a molecule.\cite{Alex,Babiker} One such tailored optical mode is a Laguerre-Guass beam. 

Laguerre-Gauss (LG) beams have revolutionized the field of optical physics since the discovery that they possess a well-defined orbital form of electromagnetic angular momentum (OAM).\cite{Allen1992} Many authors have focused on the dependence of field amplitude of LG beams on the $\phi$-coordinate in a cylindrical coordinate system since the OAM of these beams derives from this dependence.\cite{Allen1992,allen2003optical,Padgett199536} Despite intense interest in the OAM of LG beams, some authors have found that the full spatial dependence of the field of a LG beam, including its radial variation, may also lead to new, interesting effects in light-matter interactions.

First among this group was van Enk, who explored how quantization of the center of mass motion of atoms affects selection rules for light mediated atomic transitions.\cite{vanEnk_1} In discussing the coupling of the 'external' center-of-mass motion of an atomic system that is dominated by nuclear motion to a light field, van Enk points out \emph{"These changes originate from the fact that the external center-of-mass position $R$ must be considered a dynamical variable. Consequently, both the transition probability and the selection rules depend on the explicit spatial dependence of the field."} This comment was made in the context of the external motion of the nucleus of an ultracold atom. However, the distinction between 'external' nuclear motion of an atom and 'internal' nuclear motion of a poly-atomic molecule is blurry. 

The work of Alexandrescu, Cojoc, and Di Fabrizio highlights this blurry distinction between 'external' and 'internal' nature of nuclear motion in atomic versus molecular systems, respectively. Over a decade after van Enk's work, Alexandrescu \emph{et al.} investigated the interaction of a purely azimuthal LG beam with a molecular ion that possessed only a single electron, H$_2^+$.\cite{Alex} Using the Power-Zenau-Wooley interaction Hamiltonian formalism, these authors showed that the radial variation of an incident LG beam couples directly to the internal vibrational degree of freedom of this simple molecule. This coupling amends the Franck-Condon vibrational overlaps that determine which vibrational states participate in a resonant, vibronic transition. Building on this idea of the coupling of the radial variation and the internal degrees of freedom of matter, van Veenandaal and McNutly showed how an incident LG beam affects dichroism in molecular x-ray absorption processes via selective excitation of higher order electronic multipole moments of different molecules.\cite{NcNulty}

Despite the prospect of the radial variation of LG beams to provide novel changes to light-molecule interactions, published experimental results focus instead on the transfer of OAM to molecular systems. These studies investigate the linear absorption of visible light carrying OAM and LG beams by molecules.\cite{Woerdman,Araoka} However, no study has observed changes in the electronic behavior of molecules in the presence of LG beams or OAM. 

While experimental techniques taking advantage of the linear absorption of visible light can provide information related to a molecule's electronic structure, these techniques are largely insensitive to small changes in the nuclear structure of molecules. With an inherently more sensitive probing of molecular structure than linear absorption, light scattering may provide the technique of choice in which to discover the strength and utility of OAM and LG beams in the spectroscopist's toolbox in deciphering and possibly controlling nuclear dynamics and behavior central to a wide array of applications.

Despite research into the absorption of LG beams by molecules, little work investigates the scattering of LG beams from these systems. This paper investigates the inelastic scattering of classical LG beams from complex, poly-atomic molecules when the frequency of the incident light field matches a vibronic transition in the molecule. The paper is organized into four sections. After the introductory section, the second section derives the Hamiltonian that describes the interaction between a molecule and the classical LG beam, proposes a simple vibronic model molecule with which to simulate the effects of the LG-molecule interaction and uses the interaction Hamiltonian to derive the third order nonlinear response function of the molecule in the case of a LG beam. In the third section, we use the third order nonlinear response function for the molecule-LG interaction to derive an analytic form of the signal measured in a spontaneous light emission measurement excited by an incident, classical LG beam and explore the physical conditions under which Raman scattering dominates the molecule's response. This method of derivation allows us to calculate an effective differential resonance Raman scattering cross-section from which changes to selection rules are examined due to the presence of the transverse coupling between the incident light field and molecule's quantized vibrations. Deriving the third order nonlinear response of a vibronic molecule to an incident, classical LG beam also allows future theoretical studies and predictions of the effects of these novel fields in other nonlinear optical spectroscopies including \emph{stimulated} Raman scattering. We also explore how the derived results from the previous sections may be applied to resonant inelastic x-ray scattering (RIXS) and propose how the predicted effects may be visible in experiments measuring the scattering of x-rays from poly-atomic molecules. The fourth section contains a summary and concluding remarks.  

This study shows three novel facets of LG beam-molecule interactions that make their use in resonant Raman scattering desirable. First, we reproduce the coupling between the transverse radial variation of the LG beam and the vibrations of the molecule first predicted by Alexandrescu \emph{et al.}. We extend their treatment to show that this radial-coupling mechanism represents a breakdown of the adiabatic approximation directly caused by the electromagnetic interaction.\cite{harris1989symmetry,fischer1984vibronic} Electromagnetically induced breakdown of the strict separation of the electronic and nuclear coordinates leads to changes in which specific vibrational transitions are allowed upon an electronic interaction, i.e. changes in the vibronic selection rules. Therefore, the coupling between the radial variation of the incident, classical LG field and the molecule's nuclear motion represents an electromagnetically induced vibronic coupling of the moelcule's degrees of freedom.  

Second, we find that the coupling of the transverse, radial variation of a LG beam to the vibrational degrees of freedom of a polyatomic molecule does not necessitate the placement of the molecule at the center of the incident LG beam. Given the cylindrical symmetry of the incident LG beam, all molecules within the transverse profile of the mode experience the same radial variation and, therefore, the same coupling between the mode's radial variation and their vibrational degrees of freedom. This result contrasts to those of previous investigations that study the transfer of OAM from a LG beam to a molecule in which a molecule must exist at the center of the beam to 'sense' the full azimuthal variation of the transverse mode of the incident field.\cite{Alex,Babiker} 

Third, for a model vibronic molecule, we show that the coupling between the radial variation of the incident LG beam and the vibrations of the molecule enhances one of the two quantum pathways participating in a resonant inelastic scattering process. We find that situations in which the frequency of the vibration of interest approaches the linewidth of the resonant transition, the two participating vibronic transitions spectrally overlap. In such a case, the enhancement of one pathway caused by the LG beam excitation coherently and incoherently modulates the scattered intensity. This modulation is dependent on the detuning from each resonant vibronic excitation transitions, the frequency of the vibration of interest, the relative phase of the emission vibronic transition moments and magnitude of the transverse coupling constant $D$, which relates the amplitude of the vibration to the waist of the incident beam. For large values of $D$, the LG beam increases the scattering intensity when the laser is tuned to either the vibronic transition through the higher lying intermediate state or that through the lower lying intermediate state. The enhancement from the LG beam also shifts the peaks and minima in the detuning spectrum of the differential resonance Raman scattering cross-section. 

The modulation explored in this study leads to the possibility to coherently control molecular behavior using LG beams in resonant enhanced light scattering. This form of coherent control is especially important in x-ray-molecule interactions. While technologies exist to actively manipulate the spectral composition and phase of pulses in the UV, visible and IR regions of the EM spectrum, technologies to spectrally shape pulses of synchrotron x-rays do not yet exist. Some researchers have shown the ability to control the shape of soft-ray pulses produced through high harmonic generation (HHG) of shaped ultrafast pulses, but photon energies using this method do not exceed a few tens of eV.\cite{Gerber_xray_shaping} While some theoretical studies show the possibility of HHG to provide higher energy shaped pulses for coherent control, experiments have yet to confirm these predictions.\cite{HHG_shaping_theory,molecule_x-ray_shaping}  

By varying the frequency and mode of the incident field and the vibrational frequency of interest, researchers may be able control the resonant x-ray scattering intensity and spectra excited in the energy region of many core electronic transitions via the mechanisms explored in this study. Active and systematic control over the scattering intensity provides new ways to attain information about the electronic structure, dynamics and reactivity of complex molecules. By mapping out those parameters as a function of the incident laser mode, LG-mediated coherent molecular control could allow researchers to better understand the behavior of poly-atomic molecules including biomolecules, unknown molecular samples and molecular materials important to fields ranging from energy science to information technology, including nanostructures and quantum materials. 

\section{Theoretical Derivation of LG-Excited Nonlinear Optical Response Functions}
\subsection{LG-Molecule Interaction Hamiltonian}
This treatment begins with the derivation of the Hamiltonian describing the interaction between a classical, monochromatic (cw) LG beam and a molecule in the weak field limit. A LG beam solves the paraxial Helmholtz equation in cylindrical coordinates.\cite{allen2003optical,Allen1992}. The field amplitude becomes, 
\begin{eqnarray}
u_{p\ell}\left(r,\phi,z\right)=\frac{C}{\sqrt{1+z^2/z_R^2}}\left[\frac{r\sqrt{2}}{w(z)}\right]^\ell L_p^\ell\left[\frac{2r^2}{w^2(z)}\right]\nonumber\\
\times exp\left[\frac{-r^2}{w^2(z)}\right]exp\left[\frac{-\imath kr^2z}{2\left(z^2+z_R^2\right)}\right]exp\left(-\imath \ell\phi\right)\nonumber\\\times exp\left[\imath(2p+\ell+1)tan^{-1}\left(\frac{z}{z_R}\right)\right]e^{\imath \vec{k}\cdot\vec{r}}.
\end{eqnarray}
where $C$ is a normalization constant, $w(z)$ is the z-position dependent radius of the beam, $z_R$ is the Rayleigh range, the beam waist, $w_0$ is located at $z=0$ and $\vec{k}\cdot\vec{r}=kz$ for a LG beam propagating in the $z$-direction. $L_p^\ell$ represents the associated Laguerre polynomials. One should note that when $p=0$, but $\ell\neq0$ the functional form of the field amplitude simplifies significantly. For this treatment, we calculate the changes to resonant inelastic scattering of incident Laguerre-Gauss beams when the vibration of the molecule is a small fraction of the size of the waist of the incident beam. This assumption allow us to write the amplitude of the LG beam as,
\begin{eqnarray}
\vec{E}(\vec{r},t)=\mathcal{C}_\ell\vec{E}_0\mathcal{R}_{|\ell|}^\ell\left(\vec{r}\right)e^{\imath(\vec{k}\cdot\vec{r}-\omega t)}.
\end{eqnarray}
where the Gaussian roll-off of the field amplitude in Eq. (1) has been neglected.  In Eq. (2), $\mathcal{C}_\ell=(-)^{\frac{\ell+|\ell|}{2}}2^{|\ell|}\sqrt{|\ell|!}$ is the normalization constant of the field, $\mathcal{R}_\ell^m(\vec{r})=\mathcal{N}_{\ell,m}\times\left(\frac{r}{w_0}\right)^\ell Y_\ell^m(\theta,\phi)$ are the regular solid spherical harmonics and $\vec{E}_0$ is the vector amplitude of the field. Normalization of the regular spherical harmonics obeys the relation $\mathcal{N}_{\ell,m}=\sqrt{\frac{4\pi}{(2\ell+1)(\ell+m)!(\ell-m)!}}$. Figure 1 shows the region of the transverse profile of the LG beam over which the approximation made in Eq. (2) is valid.  

The derivation presented here utilizes the Power-Zienau-Wooley (PZW), or multipolar formalism, of the interaction Hamiltonian.\cite{power,craig,mukamel,cohen} Due to the functional form and physical model of the PZW canonical transformation, the field variation of a classical LG beam enters directly into the interaction Hamiltonian, as seen previously in the derivation by Alexandrescu \emph{et al.}\cite{Alex} This makes new and interesting interactions in molecular systems more transparent to the investigator and easier to interpret in resonant transitions.  

Several texts on nonrelativistic quantum electrodynamics discuss the PZW transformation and its relation to the typical $\vec{p}\cdot\vec{A}$, or minimally-coupling, interaction Hamiltonian.\cite{power,craig,mukamel,cohen} It is worth mentioning that there is nothing intrinsically special about the PZW transformed interaction Hamiltonian or the behavior explored in context of this transformation. The transformation simply allows one to expose augmented quantum molecular behavior with a greater ease, especially in the context of polyatomic molecules.\cite{mukamel} 

To begin, a polyatomic molecule is illuminated with a classical, monochromatic LG beam whose amplitude indices are generally described as $p=0, \ell\neq0$. We also assume that the molecule is located anywhere within the beam. To account for the dependence of the interaction on the position of the molecule within the transverse profile of the beam, we write the coordinates of the molecule in the reference frame of the field as, 
\begin{eqnarray}
\vec{r}_m^{field}=\vec{r}_{field}+\vec{r}_m.
\end{eqnarray}
where $r_{field}$ represents the distance from the center of the beam to the location of the molecule within the transverse radial profile of the beam for which Eq. (2) is valid. When the molecule is located at the center of the beam, we set the vector $\vec{r}_{field}=0$. 

After the PZW transformation is complete, in the absence of a magnetic field, the interaction Hamiltonian is,
\begin{eqnarray}
H_{int}=-\int\vec{E}\cdot\vec{P}d^3r.
\end{eqnarray}
where $\vec{E}$ is the incident electric field and
\begin{eqnarray}
\vec{P}=\sum_\alpha \vec{\mathcal{P}}_\alpha.
\end{eqnarray}
is the macroscopic polarization treated as a sum over effective electric dipoles summed over the the $\alpha$ microscopic polarization fields due only to the electrons of this molecule, 
\begin{eqnarray}
\vec{\mathcal{P}}_\alpha=\sum_mq_{m\alpha}\vec{r}_{m\alpha}\int_0^1du\delta\left[\vec{r}-\vec{R}_m-u\vec{r}_{m\alpha}\right].
\end{eqnarray}
\begin{figure}
\centerline{\includegraphics[width=8.5 cm]{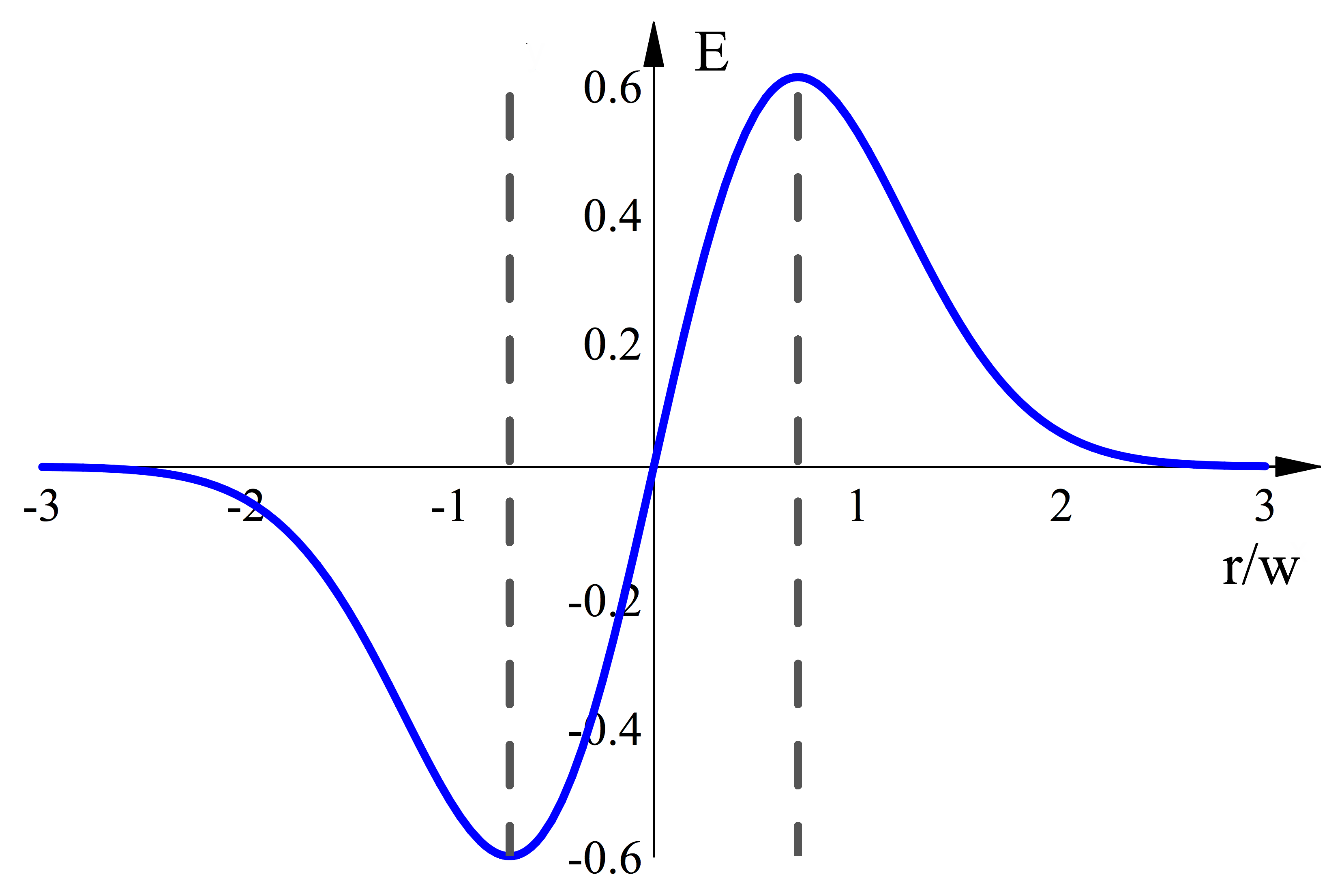}}
\caption{\label{FIG. 1}(color online) The transverse region of a low order Laguerre-Guass beam (p=0, $\ell=1$) over which the form of Eq. (2) is valid is located between the dashed lines.}
\end{figure}
In Eq. (6) we have transformed the coordinate dependence of the microscopic polarization into the center of mass reference frame of the molecule.  
From there, we transform into the field coordinate system and the microscopic polarization fields become,
\begin{eqnarray}
\vec{\mathcal{P}}_\alpha=\sum_mq_{m\alpha}\vec{r}_{m\alpha}^{field}\int_0^1du\delta\left[\vec{r}-\vec{R}^{field}_m-u\vec{r}_{m\alpha}^{field}\right],
\end{eqnarray}
where
\begin{subequations}
\begin{eqnarray}
\vec{R}_m^{field}=\vec{r}_{field}+\vec{R}_m,\\
\vec{r}_{m\alpha}^{field}=\vec{r}_{field}+\vec{r}_{m\alpha},\\
u\vec{r}_{m\alpha}^{field}=u\vec{r}_{field}+u\vec{r}_{m\alpha}.
\end{eqnarray}
\end{subequations}

Again, the subscript $field$ denotes the coordinates of the incident LG beamÕs electric field while the superscript $field$ denotes the coordinates of the molecule with reference to the center of the incident beam. The transformed macroscopic polarization becomes,
\begin{eqnarray}
\vec{P}=\sum_{\alpha,m}q_m\left(\vec{r}_{m\alpha}+\vec{r}_{field}\right),\nonumber\\\times\int_0^1du\delta\left(\vec{r}-\vec{R}_m-u\vec{r}_{m\alpha}-\vec{r}_{field}\left[1+u\right]\right).
\end{eqnarray} 

The molecule of interest is defined by an adiabatically separated wavefunction with electronic and vibrational degrees of freedom, $|\psi\rangle=|n\rangle|\nu\rangle$. The external radial spherical coordinate of $\alpha^{th}$ electron is $\vec{r}_{m\alpha}$ and the center-of-mass position in this coordinate system is $\vec{R}_m$ where $\vec{R}_m=\frac{\left(\sum_iM_i\vec{r}_{mi}\right)}{M_{tot}}$; $M_i$ is the mass of the $i^{th}$ particle and $M_{tot}$ is the total mass of the molecule. The position of one of the molecule's nuclei relative to the center-of-mass of the molecule is $\vec{r}_{nuc}$. The center of mass representation of these coordinates obey the relation $\vec{r}_i^{c.m.}=\vec{r}_i-\vec{R}$.

Using these definitions, the interaction Hamiltonian in Eq. (4) due to the $\alpha^{th}$ electron becomes, 
\begin{widetext}
\begin{eqnarray}
H_{int}=-\int_0^1\sum_{m,\alpha} q_{m\alpha}\mathcal{C}_\ell \vec{E}_0\cdot\vec{r}'_{m\alpha}\mathcal{R}_{|\ell|}^{\ell}\left(\vec{r}\right)e^{\imath(\vec{k}\cdot\vec{r}-\omega t)}du\delta\left[\vec{r}-\vec{R}_m-u\vec{r}'_{m\alpha}\right]d\vec{r}\nonumber\\
=-\sum_{m,\alpha}q_{m\alpha}\mathcal{C}_\ell\vec{E}_0\cdot\vec{r}'_{m\alpha}\int_0^1du\mathcal{R}_{|\ell|}^{\ell}\left(\vec{R}_m-u\vec{r}'_{m\alpha}\right)e^{\imath\left[\vec{k}\cdot\left(\vec{R}_m-u\vec{r}'_{m\alpha}\right)-\omega t\right]}\nonumber\\=-\sum_{m,\alpha}q_{m\alpha}\mathcal{C}_\ell\vec{E}_0\cdot\vec{r}'_{m\alpha}\sum_{\ell_1=0}^{|\ell|}\sum_{m_1=-\ell_1}^{\ell_1}(-)^{\ell_1}\mathcal{R}_{\ell_1}^{m_1}(\vec{R}_m)\mathcal{R}_{|\ell|-\ell_1}^{\ell-m_1}(\vec{r}'_{m\alpha})e^{\imath(kz_R-\omega t)}
\left[\frac{1}{|\ell|-\ell_1+1}+\frac{\imath kz_e}{|\ell|-\ell_1+2}\right]\nonumber\\
=-\sum_{m,\alpha}q_{m\alpha}\mathcal{C}_\ell\vec{E}_0\cdot\vec{r}'_{m\alpha}\sum_{\ell_1=0}^{|\ell|}\sum_{m_1=-\ell_1}^{\ell_1}(-)^{\ell_1}\mathcal{R}_{\ell_1}^{m_1}(\vec{R}_m)\sum_{\ell_2=0}^{|\ell|-\ell_1}\sum_{m_2=-\ell_2}^{\ell_2}\mathcal{R}_{\ell_2}^{m_2}(\vec{r}_{nuc})\mathcal{R}_{|\ell|-\ell_1-\ell_2}^{\ell-m_1-m_2}(\vec{r}_{in,m})\nonumber\\\times e^{\imath(kz_R-\omega t)}\left[\frac{1}{|\ell|-\ell_1+1}+\frac{\imath kz_e}{|\ell|-\ell_1+2}\right].
\end{eqnarray}
where $\vec{r}'_{m\alpha}=\vec{r}_{m\alpha}-\vec{R}_m$ and we have used the sum rules for the regular solid spherical harmonics, $\mathcal{R}_\ell^m(\vec{r}_1\pm\vec{r}_2)=\sum_{\ell_1=0}^{\ell}\sum_{m_1=-\ell_1}^{\ell_1}(\pm)^{\ell_1}\mathcal{R}_{\ell_1}^{m_2}(\vec{r}_1)\mathcal{R}_{\ell-\ell_1}^{m-m_1}(\vec{r}_2)$. We also made the simplification $exp(\imath\vec{k}\cdot\vec{r}_{m\alpha}')=exp(\imath kz_e)\approx1+\imath kz_e$. Going from the third line of Eq. (10) to the fourth line, we have replaced the center-of-mass coordinate of the $\alpha ^{th}$ electron with $\vec{r}_{m\alpha}=\vec{r}_{nuc}+\vec{r}_{in,m\alpha}$ where $\vec{r}_{nuc}$ is the center-of-mass position of one  the molecule's nuclei and $\vec{r}_{in,m}$ is the position of the $\alpha ^{th}$ electron with respect to this nucleus. Both of these positions are calculated with respect to the center of mass of the molecule. 

To examine transitions dominated by the electronic dipole of the molecule in similarity to the treatment of Alexandrescu \emph{et al.}, we take only the terms from this interaction Hamiltonian that depend linearly on $r_{in,m}$.\cite{Alex} In addition, we only consider those terms in which the polarization of the incident field couples to the $\alpha^{th}$ electron. This allows us to focus on the vibrational behavior of the molecule in the case that the field couples to its electronic transition moments. In effect, these two physical conditions mean that we only take terms from Eq. (10) containing a dot product between the internal position of the electron, $\vec{r}_{in,m}$, and incident vector field amplitude, $\vec{E}_0$ while also only examining terms in which $\ell_2=|\ell|-\ell_1$ in the second sum over the regular solid spherical harmonics. The interaction Hamiltonian now becomes, 
\begin{align}
H_{int}^{LG}=-\sum_{m,\alpha}q_{m\alpha}\vec{r}_{in,m}\cdot\vec{E}_0\mathcal{C}_\ell e^{\imath\omega t}\sum_{\ell_1=0}^{|\ell|}\sum_{m_1=-\ell_1}^{\ell_1}(-1)^{\ell_1}\mathcal{R}_{|\ell|-\ell_1}^{\ell-m_1}(\vec{r}_{nuc})\left[\frac{1}{|\ell|-\ell_1+1}+\frac{\imath kz_e}{|\ell|-\ell_1+2}\right]\nonumber\\
=-\sum_{m,\alpha}q_{m\alpha}\vec{r}_{in,m}\cdot\vec{E}_0\mathcal{C}_\ell e^{\imath\omega t}\sum_{\ell_1=0}^{|\ell|}\frac{1}{\sqrt{(2\ell_1+1)!}}\left(\frac{r_{nuc}}{w_0}\right)^{|\ell|-\ell_1}Y_{|\ell|-\ell_1}^{sgn(\ell)\cdot(|\ell|-\ell_1)}(\hat{r}_{nuc})\nonumber\\\times\left[\frac{1}{|\ell|-\ell_1+1}+\frac{\imath kz_e}{|\ell|-\ell_1+2}\right]\times\sqrt{\frac{3/4\pi}{\Gamma\left[2(|\ell|-\ell_1)+2\right]}}
\end{align}
where we have used the same physical argument to eliminate the sum over the index $m_2$ as used in other treatments of molecule-LG beam interactions and we have neglected the coupling to the center of mass motion.\cite{Alex,alex_2} 

In the treatment presented in Ref. [13], the molecule and the incident field share a common origin. An origin common to both the molecule and field allows the molecule to 'sense' the full azimuthal variation of the incident field. Therefore, for this physical situation, the term $Y_{|\ell|-\ell_1}^{sgn(\ell)\cdot(|\ell|-\ell_1)}(\hat{r}_{nuc})$ represents a transfer of the orbital angular momentum (OAM) of the incident field to the rotational state of the molecule. 

However, for the case in which the molecule does not share a common origin with the incident field, we are effectively evaluating the functional form of the $Y_{|\ell|-\ell_1}^{sgn(\ell)\cdot(|\ell|-\ell_1)}(\hat{r}_{nuc})$ term at one position of the azimuthal phase of the incident LG beam. Since we have evaluated the electric fieldÕs azimuthal dependence at one phase point, the spherical harmonic term becomes constant and is factored out of the interaction Hamiltonian. Therefore, the coupling between the incident LG beam and molecule that is dominated by the beam's OAM depends sensitively on where the molecule is located within its transverse profile.

In contrast to the OAM-mediated coupling between the incident beam and molecule, the coupling between the radial variation of the incident beam and the illuminated molecules does not depend on the azimuthal position of the molecule within the beam. Even after factoring out the term that represents the coupling of the beam's OAM to the molecule, the interaction Hamiltonian in Eq. (11) depends on the degrees of freedom of the $m^{th}$ nucleus, $\vec{r}_{nuc}$. Physically, this independence of the coupling on the molecule's position makes sense. The radial variation of the incident LG beam is symmetric about the center of the beam. Traveling radially out from the center of the beam, the variation of the transverse mode is no different in the 'doughnut' portion of the beam than in other portions of the beam for which Eq. (2) is valid. A molecule 'senses' the same radial variation at all points within this portion of the beam when the wavelength of the incident beam is not significantly bigger than the size of the molecule, as will be discussed in more detail below. Therefore, we use Eq. (9) to describe the polarization due to a molecule located anywhere within the profile of the classical LG beam such that Eq. (2) valid. 

We now use the normal coordinates of the molecule, $\{Q_j\}$, to express the position of the nucleus of interest with respect to the center of mass of the molecule, $\vec{r}_{nuc}=\bar{r}_{nuc}+\sum_jQ_j\vec{v}_j$, where $\bar{r}_{nuc}$ represents the equilibrium position of this nucleus and $\vec{v}_j$ is the vector that describes the direction and magnitude of motion of the $j^{th}$ normal coordinate. In terms of the normal coordinates, the scalar distance of the nucleus of interest to the center of mass is,
\begin{align}
r_{nuc}=\left|\bar{r}_{nuc}\right|\sqrt{1+\frac{1}{\left|\bar{r}_{nuc}\right|^2}\left|\sum_jQ_j\vec{v}_j\right|^2-2\frac{1}{\left|\bar{r}_{nuc}\right|^2}\bar{r}_{nuc}\cdot\sum_jQ_j\vec{v}_j}\nonumber\\
\approx\left|\bar{r}_{nuc}\right|\sqrt{1-2\sum_j\frac{\left|\vec{v}_j\right|}{\left|\bar{r}_{nuc}\right|}Q_j\cos\theta_{rv}}\approx\left|\bar{r}_{nuc}\right|\left(1+\sum_j\frac{\left|\vec{v}_j\right|}{\left|\bar{r}_{nuc}\right|}Q_j\right)
\end{align}
where we have made the approximation that the equilibrium distance of the nucleus of interest from the center of mass, $\left|\bar{r}_{nuc}\right|$, is significantly larger than the displacement of this nucleus due to any of the vibrations of interest, $\left|\vec{v}_j\right|$, and that the angle between these two vectors is very small, i.e. $\theta_{rv}\approx0$. The first of these approximations is the basis for the use of the harmonic oscillator model to explain the vibrational dynamics of polyatomic molecules and the quantization of the vibrational states of these systems making it physically reasonable. The second approximation is specific to the totally symmetric ring breathing vibrations of aromatic molecules proposed as model systems with which to study the effects of LG beams on light-molecule interactions. During these vibrations, nuclei of the molecule nominally move along the radial line that connects the center of mass to each nucleus. In the case of benzene, it is strictly true that $\theta_{rv}=0$, as defined above, while this angle becomes slightly larger for substituted and poly-ringed aromatic systems. However, in all cases this angle is small enough to justify the simplifications used in Eq. (12).  

Using this replacement, the interaction Hamiltonian becomes,
\begin{eqnarray}
H_{int}^{LG}=\sum_{m,\alpha}\sum_{\sigma=0,\pm1}E_\sigma\sum_{\ell_1}^{|\ell|}\mathcal{A}_{\ell,\ell_1}\left[\sum_{s=0}^{|\ell|-\ell_1}{|\ell|-\ell_1 \choose s}\left(\sum_j\frac{|\vec{v}_j|}{|\bar{r}_{nuc}|}Q_j\right)^s\right]q_{m\alpha}r_{in}Y_1^\sigma(\hat{r}).
\end{eqnarray}
where $\vec{r}_{in,m}\cdot\vec{E}_0=r_{in,m}\sqrt{4\pi/3}\sum_{\sigma=0,\pm1}E_\sigma Y_1^\sigma(\hat{r})$, $\pm E_{\sigma}=1\sqrt{2}(E_x\pm\imath E_y),$ the case of $\sigma=0$ indicates linear polarization of the incident electric field in either the $x$ or $y$ directions, the dependence on the normal coordinate, $Q$, is re-written as a binomial expansion and
\begin{eqnarray}
\mathcal{A}_{\ell,\ell_1}=\mathcal{C}_\ell e^{\imath\omega t}\frac{4\pi}{\sqrt{3\left[(2\ell_1+1)!\right]}}\left(\frac{|\bar{r}_{nuc}|}{w_0}\right)^{|\ell|-\ell_1}\nonumber\\\times\left[\frac{1}{|\ell|-\ell_1+1}+\frac{\imath kz_e}{|\ell|-\ell_1+2}\right]\times\sqrt{\frac{3/4\pi}{\Gamma\left[2(|\ell|-\ell_1)+2\right]}}.
\end{eqnarray}
where we have retained the first two terms in the Taylor series expansion of $exp(-\imath kz_e)$ in order not to explicitly invoke the dipole approximation.

Eq. (13) shows a functional form dissimilar to the interaction of a vibronic, polyatomic molecule with a plane electromagnetic wave. Most notable is the appearance of the sum of the of molecule's normal coordinates, $\sum_jQ_j$, directly into the interaction Hamiltonian. Typically the appearance of normal coordinates in light-molecule interactions occurs due to coupling between electronic and vibrational degrees of freedom of the molecule that provides higher-order perturbative corrections to the interaction: vibronic coupling as explained above. In the case of a classical LG beam, this coupling appears as a lowest order perturbative effect due to the interaction of the incident field and the molecule: an electromagnetic induced vibronic coupling. The importance of this change in vibronic coupling is more thoroughly examined below.  

This interaction Hamiltonian is used to construct an interaction operator, V. The matrix elements of this interaction operator become,
\begin{align}
V_{\psi\psi'}=\langle\psi|H_{int}|\psi'\rangle,\nonumber\\
=\langle\psi|\sum_{m,\alpha}\sum_{\sigma=0,\pm1}E_\sigma\sum_{\ell_1}^{|\ell|}\mathcal{A}_{\ell,\ell_1}\left[\sum_{s=0}^{|\ell|-\ell_1}{|\ell|-\ell_1 \choose s}\left(\sum_j\frac{|\vec{v}_j|}{|\bar{r}_{nuc}|}Q_j\right)^s\right]q_{m\alpha}r_{in}Y_1^\sigma(\hat{r})|\psi'\rangle\nonumber\\=\sum_{m,\alpha}\sum_{\sigma=0,\pm1}\sum_{\ell_1}^{|\ell|}\mathcal{A}_{\ell,\ell_1}\langle\nu|\left[\sum_{s=0}^{|\ell|-\ell_1}{|\ell|-\ell_1 \choose s}\left(\sum_j\frac{|\vec{v}_j|}{|\bar{r}_{nuc}|}Q_j\right)^s\right]|\nu'\rangle\langle e|E_\sigma q_{m\alpha}r_{in}Y_1^\sigma(\hat{r})|e'\rangle.
\end{align}
\end{widetext}
where the electronic and vibrational components of the molecule's wavefunction, $|\psi\rangle$, have already been defined.

Before we use this interaction operator to calculate the third order nonlinear optical response of a molecule to an incident LG beam, we present the model structure of a vibronic molecule we later use to simulate the effects of the transverse coupling between the incident field and this molecule's vibration on the differential resonantly-enhanced inelastic scattering cross-section.

\subsection{Model Vibronic Molecule}
For many poly-atomic molecules, inelastic scattering processes that occur simultaneously with resonant electron transitions are allowed for both totally and non-totally symmetric vibrations. Physically, this facet of molecule-light interactions occurs because the excited electronic states of diatomic and poly-atomic molecules are displaced with respect to the equilibrium position of the ground electronic state. This displacement necessitates the use of two separate basis sets to describe the vibrational sub-levels of each respective electronic state of the molecule and relaxes strict vibronic selection rules governing transitions between quantum states of the harmonic potential well. It is the relaxation of these selection rules that allows significant enhancement of resonant inelastic scattering from molecules either for simple bond lengthening upon a resonant transition or for more complexes changes of the excited electronic state such as Jahn-Teller distortions, for example.\cite{Zink_review,albrecht}

Based on these considerations, we treat the resonant scattering of classical LG beams due to the vibrations of a model vibronic molecule. This vibronic model molecule possesses two electronic states, ground and excited, that we treat as harmonic potentials. To account for resonant electronic transitions, we denote the vibronic states $|a\rangle$ and $|c\rangle$ as states on the ground electronic manifold of vibrational states,$|g\rangle$, and $|b\rangle$ and $|d\rangle$ as states on the excited electronic manifold of vibrational states, $|e\rangle$. The excited state electronic potential energy surface is displaced from the equilibrium geometry of the ground state by an amount $d$. We assume that the frequency of the vibration along which the excited electronic state is displaced are the same in each respective electronic state and is denoted $\omega_\nu$. Figure 2 shows this vibronic model. We assume this multilevel vibronic model molecule couples to both its surrounding bath and the vacuum electromagnetic field. A model of this form has been used to explain coupling between resonant core electronic transitions and both totally and non-totally symmetric vibrations, such as those examined below.\cite{Skytt_PRA_1997,minkov:5733}    
\begin{figure}
\centerline{\includegraphics[width=8.5 cm]{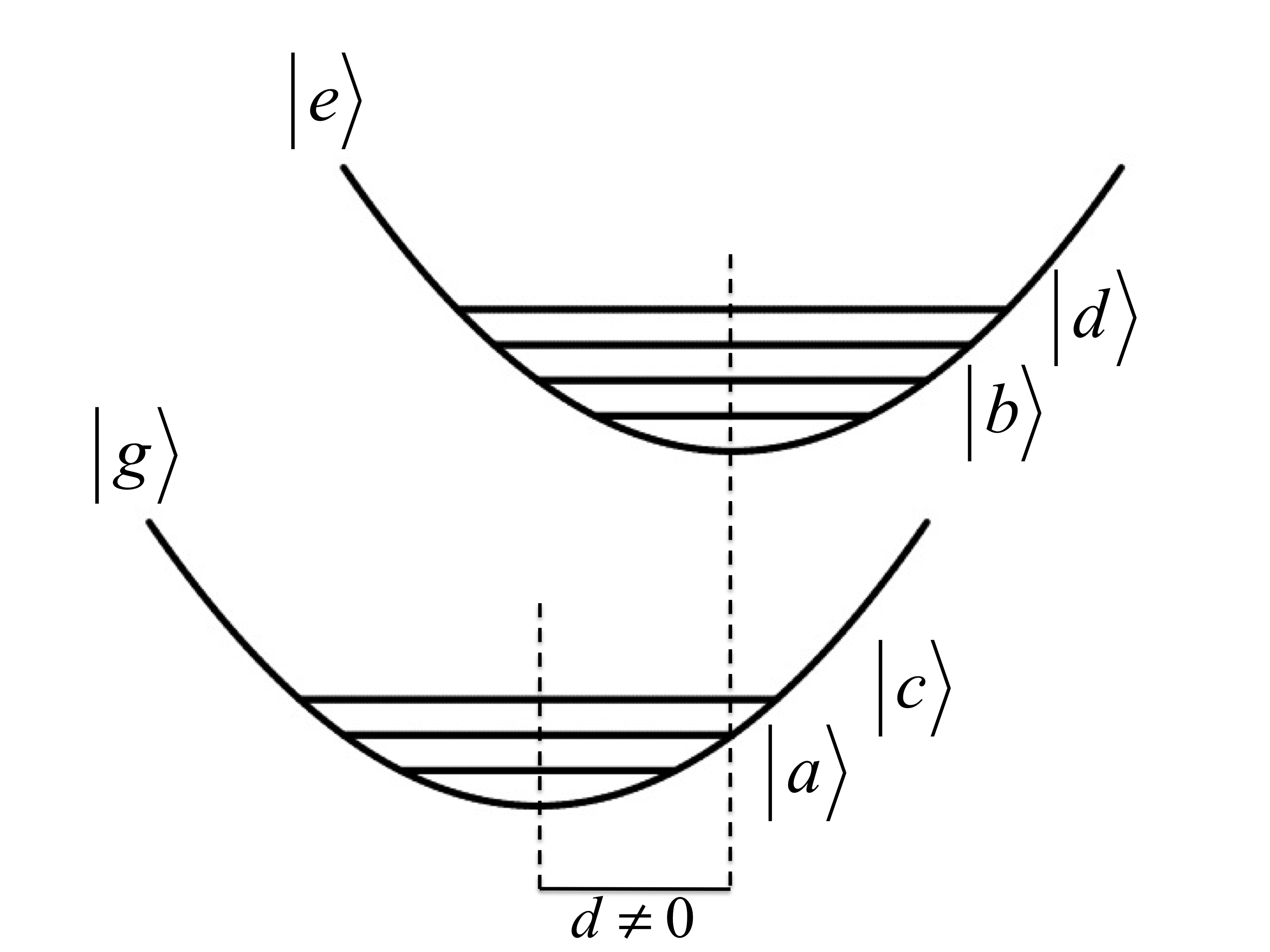}}
\caption{\label{FIG. 2} Vibronic states that contribute to the third order nonlinear response of this molecular model. The states $|a\rangle$ and $|c\rangle$ exist on the ground electronic state, $|g\rangle$, while the states $|b\rangle$ and $|d\rangle$ exist on the excited electronic state,$|e\rangle$.}
\end{figure}

For this treatment of resonant inelastic light scattering, we assume the interaction operator, $V$, derived in the previous section can only connect vibronic states on different electronic manifolds. Using Fig. 2 as a guide, we see that the interaction operator connects the state $|a\rangle$ to the state $|b\rangle$ via an excitation process, the state $|d\rangle$ to the state $|a\rangle$ through an emission process, the state $|c\rangle$ to the state $|d\rangle$ through an excitation process and the state $|b\rangle$ to the state $|c\rangle$ through an emission process. This collection of transitions accounts for both spontaneous Raman and fluorescence processes in molecules well-approximated by this vibronic model system.\cite{mukamel} 

The intensity of resonant, inelastic scattering of light by molecules depends sensitively on the displacement of the excited electronic state with respect to the equilibrium position of the ground electronic state along a vibration of interest. To simulate the effect of the displacement on the spontaneous inelastic scattering of classical LG beams, we present analytical forms of the vibrational overlaps present in the matrix elements of the interaction Hamiltonian in the basis of the vibronic molecular quantum states. In particular, we calculate the overlaps in the cases that $|\psi\rangle=|n\rangle|\nu\rangle$ corresponds to the states $|a\rangle=|g\rangle|0\rangle$, $|c\rangle=|g\rangle|1\rangle$, $|b_1\rangle=|e\rangle|0\rangle$, and $|b_2\rangle=|e\rangle|1\rangle$. These matrix elements depend on explicit Franck-Condon overlap integrals through the relations, 
\begin{subequations}
\begin{align}
\langle a|H_{int}^{LG}|b_1\rangle\propto|\langle 0|0\rangle|+|\langle 0|Q|0\rangle|=\left(1+\bar{d}\right)e^{-\bar{d}^2/2},\\
\langle a|H_{int}^{LG}|b_2\rangle\propto|\langle 0|1\rangle|+|\langle 0|Q|1\rangle|=\left(1+\bar{d}\right)e^{-\bar{d}^2/2},\\
\langle c|H_{int}^{LG}|b_1\rangle\propto|\langle 1|0\rangle|=\bar{d}e^{-\bar{d}^2/2},\\
\langle c|H_{int}^{LG}|b_2\rangle\propto|\langle 1|1\rangle|=e^{-\bar{d}^2/2}.
\end{align}
\end{subequations}
where we have assumed that $|\langle 0|Q|1\rangle|=|\langle 0|0\rangle|$ and $|\langle 0|Q|0\rangle|=|\langle 0|1\rangle|$, $\bar{d}=d\sqrt{m_e\omega_\nu/2\hbar}$ is the dimensionless displacement of the excited electronic potential and we are denoting the real part of these matrix elements with square brackets.\cite{Condon,harris1989symmetry} Most noticeable about this set of equations are Eqs. (16a) and (16b). While the other overlap integrals depend on only one vibrational overlap, Eqs. (16a) and (16b) depend on two. The presence of the second vibrational overlap in the functional form of these excitation matrix elements is an important consequence of the excitation induced by the classical LG beam in this investigation and was first exposed by Alexandrescu \emph{et al.}\cite{Alex} 

We see that the radial variation of the LG beam embeds a quantum of the molecule's normal coordinate, $Q$, into the interaction Hamiltonian, as seen in Eq. (13). Embedding $Q$ in these matrix elements directly affects the magnitude of the overlap between vibrational contributions to the vibronic wavefuntions $|a\rangle$ and each state, $|b_1\rangle$ and $|b_2\rangle$. By embedding the normal coordinates into the interaction Hamiltonian, the LG beam mediates a coupling between the resonant electronic transition and nuclear states of the molecule. In this sense, resonant excitation with the classical LG beam results in a vibronic coupling as stated above. 

To understand the effects of this electromagnetically-mediated vibronic coupling on inelastic scattering, we derive the third order nonlinear response function of the molecule in the frequency domain. From this response function we can treat the situations in which resonant Raman scattering dominate the spontaneous emission from an excited molecule to expose how the functional, radial variation of the incident LG beam affects the scattering process.   

While we are interested in only the case of spontaneous scattering excited by a LG beam, we will use the interaction Hamiltonian derived in the previous section to calculate the frequency domain third order nonlinear optical response function that accounts for both spontaneous and stimulated scattering. One can then use this response function in future treatments to examine the role of transverse coupling between incident LG fields and molecules in a scenario with two LG beams for the case of stimulated scattering. 

\subsection{Nonlinear Optical Response due to LG Beam}
Using the vibronic model molecule explored in the previous section, the third order nonlinear response for three incident laser beams of frequency $\omega_1$, $\omega_2$, and $\omega_3$ is 
\begin{align}
S^{(3)}\left(\omega_1+\omega_2+\omega_3,\omega_1+\omega_2,\omega_1\right)
=\left(\frac{-1}{\hbar}\right)^3\nonumber\\\times\langle\langle V|\mathcal{G}\left(\omega_1+\omega_2+\omega_3\right)\mathcal{V}\mathcal{G}\left(\omega_1+\omega_2\right)\mathcal{V}\mathcal{G}\left(\omega_1\right)\mathcal{V}|\rho(-\infty)\rangle\rangle.
\end{align}
where 
\begin{align}
\mathcal{G}(\omega_i)=\sum_{\psi,\psi'}|\psi\psi'\rangle\rangle\mathcal{G}_{\psi\psi'}(\omega_i)\langle\langle\psi\psi'|.
\end{align}
is the frequency domain Green's function when it is projected into the vector space of the density matrix of the multilevel vibronic molecule. Since we are interested in resonant scattering processes, this Green's function is taken in the rotating wave approximation (RWA) and includes population relaxation and dephasing processes phenomenologically as
\begin{align}
\mathcal{G}_{\psi\psi'}(\omega_i)=\frac{1}{\omega_i-\omega_{\psi\psi'}+\imath\Gamma_{\psi\psi'}}.
\end{align}
$\omega_i$ is the frequency of the incident field that causes this Green's function, $\omega_{\psi\psi'}$ is the frequency of the transition between the vibronic states $|\psi\rangle$ and $|\psi'\rangle$ of the molecule and $\Gamma_{\psi\psi'}$ is the phenomenological dephasing rate of their vibronic coherence. $\Gamma_{\psi\psi'}=\gamma_{\psi\psi'}/2+\hat{\Gamma}_{\psi\psi'}$ where $\gamma_{\psi\psi'}$ is the population relaxation rate from the state $|\psi\rangle$ to the state $|\psi'\rangle$ and $\hat{\Gamma}_{\psi\psi'}$ is the pure dephasing rate connecting these same two states.\cite{mukamel}  

In the basis of the vibronic levels of the molecule, this interaction operator becomes, 
\begin{align}
V=\sum_{\substack{\psi=a,c\\\psi'=b,d}}\left[V_{\psi\psi'}|\psi\rangle\langle \psi'|+V_{\psi'\psi}|\psi'\rangle\langle \psi|\right],\nonumber\\=\sum_{\substack{a,c\\b,d}}\left[V_{ab}|a\rangle\langle b|+V_{da}|d\rangle\langle a|+V_{cd}|c\rangle\langle d|+V_{bc}|b\rangle\langle c|\right].
\end{align}
where we have defined the functional form of the interaction operator from the interaction Hamiltonian in Eq. (15).

We evaluate Eq. (17) in steps from right to left. The first operation is $\mathcal{V}|\rho(-\infty)\rangle\rangle$, corresponds to the interaction of a cw LG beam of frequency $\omega_1$ with the vibronic molecule in its initial state of the density matrix, which we denote $|aa\rangle\rangle=|a\rangle\langle a|$. This means the molecule begins as a ground state population as would be expected for room and low temperature core vibronic excitations explored in Section \Rmnum{3}. To begin,  
\begin{align}
\mathcal{V}|aa\rangle\rangle=\left[V,|a\rangle\langle a|\right]=V|a\rangle\langle a|-|a\rangle\langle a|V\nonumber\\=V_{da}|d\rangle\langle a|-V_{ab}|a\rangle\langle b|=\rho_1.
\end{align}
Following this first interaction, the molecule evolves in frequency space by interacting with its environment. Physically this means that the coherences produced by the first field interaction dephase due to interactions with the surrounding thermal bath or the fluctuations of the vacuum electromagnetic field, as expected in spontaneous light scattering. This evolution is captured by the Green's function $\mathcal{G}\left(\omega_1\right)$. The density matrix becomes,
\begin{widetext}
\begin{align}
\mathcal{G}\left(\omega_1\right)|\rho_1\rangle\rangle=\sum_{c',d'}|c'd'\rangle\rangle\mathcal{G}_{c'd'}\left(\omega_1\right)\langle\langle c'd'|\rho_1\rangle\rangle=\sum_{c',d'}\left[V_{da}\mathcal{G}_{c'd'}\left(\omega_1\right)\langle\langle c'd'|da\rangle\rangle|c'd'\rangle\rangle-V_{ab}\mathcal{G}_{c'd'}\left(\omega_1\right)\langle\langle c'd'|ab\rangle\rangle|c'd'\rangle\rangle\right]\nonumber\\=\sum_{c',d'}\left[V_{da}\mathcal{G}_{c'd'}\left(\omega_1\right)\delta_{c'd}\delta_{d'a}|c'd'\rangle\rangle-V_{ab}\mathcal{G}_{c'd'}\left(\omega_1\right)\delta_{c'a}\delta_{d'b}|c'd'\rangle\rangle\right]\nonumber\\=V_{da}\mathcal{G}_{da}\left(\omega_1\right)|d\rangle\langle a|-V_{ab}\mathcal{G}_{ab}\left(\omega_1\right)|a\rangle\langle b|=V_{da}\mathcal{G}_{da}\left(\omega_1\right)|da\rangle\rangle-V_{ab}\mathcal{G}_{ab}\left(\omega_1\right)|ab\rangle\rangle=|\rho_2\rangle\rangle.
\end{align} 

The next interaction term then acts on $\rho_2$ to produce,
\begin{align}
\mathcal{V}|\rho_2\rangle\rangle=\left[V,\rho_2\right]=V\rho_2-\rho_2V\nonumber\\=V_{da}\mathcal{G}_{da}\left(\omega_1\right)V|d\rangle\langle a|-V_{da}\mathcal{G}_{da}\left(\omega_1\right)|d\rangle\langle a|V-V_{ab}\mathcal{G}_{ab}\left(\omega_1\right)V|a\rangle\langle b|+V_{ab}\mathcal{G}_{ab}\left(\omega_1\right)|a\rangle\langle b|V\nonumber\\
=V_{cd}V_{da}\mathcal{G}_{da}\left(\omega_1\right)|c\rangle\langle a|-V_{ab}V_{da}\mathcal{G}_{ab}\left(\omega_1\right)|d\rangle\langle b|-V_{ab}V_{da}\mathcal{G}_{da}\left(\omega_1\right)|d\rangle\langle b|+V_{ab}V_{bc}\mathcal{G}_{ab}\left(\omega_1\right)|a\rangle\langle c|\nonumber\\=V_{cd}V_{da}\mathcal{G}_{da}\left(\omega_1\right)|ca\rangle\rangle-V_{ab}V_{da}\mathcal{G}_{ab}\left(\omega_1\right)|db\rangle\rangle-V_{ab}V_{da}\mathcal{G}_{da}\left(\omega_1\right)|db\rangle\rangle+V_{ab}V_{bc}\mathcal{G}_{ab}\left(\omega_1\right)|ac\rangle\rangle\nonumber\\=|\rho_3\rangle\rangle.
\end{align}

We see that this interaction produces coherences between vibronic states on the same electronic state. When $|a\rangle=|c\rangle$ or $|b\rangle=|d\rangle$, this density matrix describes a vibronic population, as further examined below. The next Green's function describes dissipation of this density matrix,
\begin{align}
\mathcal{G}\left(\omega_1+\omega_2\right)|\rho_3\rangle\rangle=\sum_{c'd'}|c'd'\rangle\rangle\mathcal{G}_{c'd'}\left(\omega_1+\omega_2\right)\langle\langle c'd'|\rho_3\rangle\rangle\nonumber\\=\sum_{c'd'}\left.[V_{cd}V_{da}\mathcal{G}_{da}\left(\omega_1\right)\mathcal{G}_{c'd'}\left(\omega_1+\omega_2\right)\langle\langle c'd'|ca\rangle\rangle|c'd'\rangle\rangle\right.\nonumber\\\left.-V_{ab}V_{da}\mathcal{G}_{ab}\left(\omega_1\right)\mathcal{G}_{c'd'}\left(\omega_1+\omega_2\right)\langle\langle c'd|db\rangle\rangle|c'd'\rangle\rangle-V_{ab}V_{da}\mathcal{G}_{da}\left(\omega_1\right)\mathcal{G}_{c'd'}\left(\omega_1+\omega_2\right)\langle\langle c'd|db\rangle\rangle|c'd'\rangle\rangle\right.\nonumber\\\left.+V_{ab}V_{bc}\mathcal{G}_{ab}\left(\omega_1\right)\mathcal{G}_{c'd'}\left(\omega_1+\omega_2\right)\langle\langle c'd|ac\rangle\rangle|c'd'\rangle\rangle\right.]\nonumber\\=V_{cd}V_{da}\mathcal{G}_{da}\left(\omega_1\right)\mathcal{G}_{ca}\left(\omega_1+\omega_2\right)|ca\rangle\rangle-V_{ab}V_{da}\mathcal{G}_{ab}\left(\omega_1\right)\mathcal{G}_{db}\left(\omega_1+\omega_2\right)|db\rangle\rangle\nonumber\\-V_{ab}V_{da}\mathcal{G}_{da}\left(\omega_1\right)\mathcal{G}_{db}\left(\omega_1+\omega_2\right)|db\rangle\rangle+V_{ab}V_{bc}\mathcal{G}_{ab}\left(\omega_1\right)\mathcal{G}_{ac}\left(\omega_1+\omega_2\right)|ac\rangle\rangle\nonumber\\=|\rho_4\rangle\rangle.
\end{align}
The next field interaction produces the density matrix,
\begin{align}
\mathcal{V}|\rho_4\rangle\rangle=\left[V,\rho_4\right]=V\rho_4-\rho_4V\nonumber\\=V_{cd}V_{da}\mathcal{G}_{da}\left(\omega_1\right)\mathcal{G}_{ca}\left(\omega_1+\omega_2\right)V|c\rangle\langle a|-V_{ab}V_{da}\mathcal{G}_{ab}\left(\omega_1\right)\mathcal{G}_{db}\left(\omega_1+\omega_2\right)V|d\rangle\langle b|\nonumber\\-V_{ab}V_{da}\mathcal{G}_{da}\left(\omega_1\right)\mathcal{G}_{db}\left(\omega_1+\omega_2\right)V|d\rangle\langle b|+V_{ab}V_{bc}\mathcal{G}_{ab}\left(\omega_1\right)\mathcal{G}_{db}\left(\omega_1+\omega_2\right)V|a\rangle\langle c|\nonumber\\-V_{cd}V_{da}\mathcal{G}_{da}\left(\omega_1\right)\mathcal{G}_{ca}\left(\omega_1+\omega_2\right)|c\rangle\langle a|V+V_{ab}V_{da}\mathcal{G}_{ab}\left(\omega_1\right)\mathcal{G}_{db}\left(\omega_1+\omega_2\right)|d\rangle\langle b|V\nonumber\\+V_{ab}V_{da}\mathcal{G}_{da}\left(\omega_1\right)\mathcal{G}_{db}\left(\omega_1+\omega_2\right)|d\rangle\langle b|V-V_{ab}V_{bc}\mathcal{G}_{ab}\left(\omega_1\right)\mathcal{G}_{db}\left(\omega_1+\omega_2\right)|a\rangle\langle c|V\nonumber\\=V_{bc}V_{cd}V_{da}\mathcal{G}_{da}\left(\omega_1\right)\mathcal{G}_{ca}\left(\omega_1+\omega_2\right)|b\rangle\langle a|-V_{ab}V_{cd}V_{da}\mathcal{G}_{ab}\left(\omega_1\right)\mathcal{G}_{db}\left(\omega_1+\omega_2\right)|c\rangle\langle b|\nonumber\\-V_{ab}V_{cd}V_{da}\mathcal{G}_{da}\left(\omega_1\right)\mathcal{G}_{db}\left(\omega_1+\omega_2\right)|c\rangle\langle b|+V_{ab}V_{bc}V_{da}\mathcal{G}_{ab}\left(\omega_1\right)\mathcal{G}_{db}\left(\omega_1+\omega_2\right)|d\rangle\langle c|\nonumber\\-V_{ab}V_{cd}V_{da}\mathcal{G}_{da}\left(\omega_1\right)\mathcal{G}_{ca}\left(\omega_1+\omega_2\right)|c\rangle\langle b|+V_{ab}V_{bc}V_{da}\mathcal{G}_{ab}\left(\omega_1\right)\mathcal{G}_{db}\left(\omega_1+\omega_2\right)|d\rangle\langle c|\nonumber\\+V_{ab}V_{bc}V_{da}\mathcal{G}_{da}\left(\omega_1\right)\mathcal{G}_{db}\left(\omega_1+\omega_2\right)|d\rangle\langle c|-V_{ab}V_{bc}V_{cd}\mathcal{G}_{ab}\left(\omega_1\right)\mathcal{G}_{db}\left(\omega_1+\omega_2\right)|a\rangle\langle d|\nonumber\\=|\rho_5\rangle\rangle.
\end{align}

We now act with the last Green's function, $\mathcal{G}(\omega_1+\omega_2+\omega_3)$ using the same method as in Eqs. (22) and (24) to produce,
\begin{align}
\mathcal{G}(\omega_1+\omega_2+\omega_3)|\rho_5\rangle\rangle=\mathcal{G}^{(3)}|\rho_5\rangle\rangle\nonumber\\
=V_{bc}V_{cd}V_{da}\mathcal{G}^{(1)}_{da}\mathcal{G}^{(2)}_{ca}\mathcal{G}^{(3)}_{ba}|b\rangle\langle a|-V_{ab}V_{cd}V_{da}\mathcal{G}^{(1)}_{ab}\mathcal{G}^{(2)}_{db}\mathcal{G}^{(3)}_{cb}|c\rangle\langle b|\nonumber\\-V_{ab}V_{cd}V_{da}\mathcal{G}^{(1)}_{da}\mathcal{G}^{(2)}_{db}\mathcal{G}^{(3)}_{cb}|c\rangle\langle b|+V_{ab}V_{bc}V_{da}\mathcal{G}^{(1)}_{ab}\mathcal{G}^{(2)}_{db}\mathcal{G}^{(3)}_{dc}|d\rangle\langle c|\nonumber\\-V_{ab}V_{cd}V_{da}\mathcal{G}^{(1)}_{da}\mathcal{G}^{(2)}_{ca}\mathcal{G}^{(3)}_{cb}|c\rangle\langle b|+V_{ab}V_{bc}V_{da}\mathcal{G}^{(1)}_{ab}\mathcal{G}^{(2)}_{db}\mathcal{G}^{(3)}_{dc}|d\rangle\langle c|\nonumber\\+V_{ab}V_{bc}V_{da}\mathcal{G}^{(1)}_{da}\mathcal{G}^{(2)}_{db}\mathcal{G}^{(3)}_{dc}|d\rangle\langle c|-V_{ab}V_{bc}V_{cd}\mathcal{G}^{(1)}_{ab}\mathcal{G}^{(2)}_{db}\mathcal{G}^{(3)}_{ad}|a\rangle\langle d|\nonumber\\=|\rho_6\rangle\rangle.
\end{align}
where we now use superscripts to denote the frequency of the field-free evolution of the system. The superscripts (1), (2) and (3) represent frequencies of $\omega_1$, $\omega_1+\omega_2$ and $\omega_1+\omega_2+\omega_3$, respectively. The third order optical response of the system is completely calculated by operating on $|\rho_6\rangle\rangle$ with the Hilbert space interaction operator, V, and tracing over the resulting matrix. We find, 
\begin{align}
S^{(3)}\left(\omega_1+\omega_2+\omega_3,\omega_1+\omega_2,\omega_1\right)=\left(\frac{-1}{\hbar}\right)^3\sum_{\substack{a,c\\b,d}}V_{ab}V_{bc}V_{cd}V_{da}\nonumber\\\times\left[\mathcal{G}^{(1)}_{da}\mathcal{G}^{(2)}_{ca}\mathcal{G}^{(3)}_{ba}-\mathcal{G}^{(1)}_{ab}\mathcal{G}^{(2)}_{db}\mathcal{G}^{(3)}_{cb}-\mathcal{G}^{(1)}_{da}\mathcal{G}^{(2)}_{db}\mathcal{G}^{(3)}_{cb}+\mathcal{G}^{(1)}_{ab}\mathcal{G}^{(2)}_{db}\mathcal{G}^{(3)}_{dc}\right.\nonumber\\\left.-\mathcal{G}^{(1)}_{da}\mathcal{G}^{(2)}_{ca}\mathcal{G}^{(3)}_{cb}+\mathcal{G}^{(1)}_{ab}\mathcal{G}^{(2)}_{db}\mathcal{G}^{(3)}_{dc}+\mathcal{G}^{(1)}_{da}\mathcal{G}^{(2)}_{db}\mathcal{G}^{(3)}_{dc}-\mathcal{G}^{(1)}_{ab}\mathcal{G}^{(2)}_{db}\mathcal{G}^{(3)}_{ad}\right].
\end{align}
\end{widetext}
In the next section we evaluate this nonlinear optical response function in the case that the frequencies $\omega_L$ of the incident laser field and $\omega_S$  satisfy the spontaneous Raman condition: $\omega_L=\omega_S+\omega_{ac}$, where $\hbar\omega_{ac}$ is the difference in energy of the vibronic states $|a\rangle$ and $|c\rangle$ on the ground electronic manifold and $\omega_S$ is the frequency of the emitted light. This technique allows us to determine how and why the incident LG beam changes the nature of this spontaneous light scattering process. 

One should note that nothing intrinsic to the nature of the excitation beam enters into this derivation of the third order nonlinear response function of the model vibronic molecule. The form of the response function in Eq. (27) would be the same for the both the case of plane wave and LG beam excitation. However, since the interaction operator, $V$, is different for the case of an incident LG beam from that of plane wave excitation, each case leads to a different optical response. We consider these differences when applied to spontaneous scattering in the next section. 

\section{Spontaneous Inelastic Scattering of LG Beams from Vibronic Molecules}
With the third order nonlinear response function derived in the previous section [Eq. (27)], we now begin to derive the spontaneous light emission (SLE) cross-section of the model vibronic molecule in the case of an incident LG beam. We will then examine the implications of the incident LG beam in the context of Raman processes for which the pure dephasing rate, $\hat{\Gamma}$, is zero. This treatment relies heavily on the presentation of Ref. [27] and references therein. 

To begin, the scattered differential power in the range of frequencies between $\omega_S$ and $\omega_S+d\omega$ is, 
\begin{align}
I_S(\omega_L,\omega_S)=I_0(\omega_L)z\rho_0\sigma(\omega_L,\omega_S).
\end{align}
where $z$ is the length of the macroscopic sample, $\rho_0$ is the density of scattering centers and $\sigma(\omega_L,\omega_S)$ is the differential scattering cross-section of the incident classical light field from the molecule in the same frequency range. This differential scattering cross-section is connected to the response function calculated in the previous section through the relation,
\begin{align}
\sigma(\omega_L,\omega_S)=\frac{4\omega_L\omega_S^3}{9\hbar^2c^4}S_{SLE}(\omega_L,\omega_S).
\end{align}
where $S_{SLE}(\omega_L,\omega_S)$ is the optical response function that comes from Eq. (27) in the case of an incident, monochromatic light field and a spontaneously scattered field with frequencies of $\omega_L$ and $\omega_S$, respectively. We use a physical argument to understand which Green's functions in Eq. (27) will contribute to the spontaneous light emission. First, using the basis states of the model vibronic molecule introduced in Section \Rmnum{2}.B, we only take terms from Eq. (27) in which the state $|a\rangle$ couples to the excited electronic vibronic state $|b\rangle$ via an excitation due to the incident field. Second, we only keep terms of Eq. (27) whose third Green's function [$\mathcal{G}^{(3)}$] couple either $|b\rangle$ or $|d\rangle$ on the excited electronic potential to $|c\rangle$ on the ground electronic potential through the emission of a field with frequency $\omega_S$. Using these two rules, we find that the spontaneous light emission optical response function becomes, 
\begin{align}
S_{SLE}(\omega_L,\omega_S)\nonumber\\=2 Im \sum_{\substack{a,c\\b,d}}V_{ab}V_{bc}V_{cd}V_{da}\left[\langle\mathcal{G}_{cb}(-\omega_S)\mathcal{G}_{db}(0)\mathcal{G}_{ab}(-\omega_L)\rho_g\rangle\right.\nonumber\\\left.+\langle\mathcal{G}_{dc}(\omega_S)\mathcal{G}_{db}(0)\mathcal{G}_{ab}(-\omega_L)\rho_g\rangle\right.\nonumber\\\left.+\langle\mathcal{G}_{dc}(\omega_S)\mathcal{G}_{ac}(\omega_S-\omega_L)\mathcal{G}_{ab}(-\omega_L)\rho_g\rangle\right].
\end{align}
where the brackets $\langle\rangle$ indicate tracing over the states of the bath surrounding the molecule and $|\rho_g\rangle=|aa\rangle\rangle$ is the ground state density matrix, as stated above in Section \Rmnum{2}.C. 

Closer inspection of the Green's functions present in Eq. (30) allows one to identify which terms lead to fulfillment of the Raman condition, $\omega_L-\omega_S=\omega_{ac}$. Only the third term within the brackets of Eq. (30) possesses a Green's function whose value increases substantially when fulfilling this condition. In fact, one finds that this Green's function obeys an important equation:
\begin{align}
\mathcal{G}_{ac}(\omega_S-\omega_L)=\frac{1}{\omega_{ac}+\omega_S-\omega_L+\imath\Gamma}\nonumber\\=\mathcal{PP}\left(\frac{1}{\omega_{ac}+\omega_S-\omega_L}\right)-\imath\pi\delta(\omega_{ac}+\omega_S-\omega_L).
\end{align}
where $\mathcal{PP}$ represents the principal part of this fraction. Using this relation, we find that in the case that $|d\rangle=|b\rangle$, the spontaneous light emission optical response function can be separated into components that lead to Raman scattering and fluorescence. That is, the optical response function becomes,
\begin{align}
S_{SLE}(\omega_L,\omega_S)=S_{Raman}(\omega_L,\omega_S)+S_{FL}(\omega_L,\omega_S).
\end{align}
where $S_{Raman}(\omega_L,\omega_S)$ and and $S_{FL}(\omega_L,\omega_S)$ are the Raman and the fluorescence contributions to SLE optical response function, respectively.

As shown in Ref. [27], when the pure dephasing rate is zero ($\hat{\Gamma}=0$), $S_{FL}(\omega_L,\omega_S)$ becomes zero and the resonantly-enhanced Raman scattering dominates spontaneous light emission response function. Under this condition and in the case that we examine a single initial state ($|a\rangle$) and a single final state ($|c\rangle$), the differential scattering cross-section of the incident, classical LG beam from this model molecule becomes, 
\begin{align}
\sigma(\omega_L,\omega_S)\nonumber\\=\frac{8\pi\omega_L\omega_S^3}{9\hbar^2c^4}\sum_{b}\left|\frac{V_{ab}V_{bc}}{\omega_{ba}-\omega_L+\imath\Gamma}\right|^2\delta(\omega_{ac}+\omega_S-\omega_L).
\end{align}
where we only sum over the intermediate states, $|b\rangle$, and now $\Gamma=\gamma/2$. $\gamma$ is the population relaxation rate of the state $|b\rangle$ and is the inverse lifetime of this state. Eq. (33) corresponds to the Kramers-Heisenberg equation typically derived via second order perturbations to the molecular wavefunction due to the incident field. 

Despite the rigor of the derivation of Eq. (33), one must take great care in interpreting the physics this model implies for resonantly-enhanced spontaneous Raman scattering. In the spontaneous scattering process explored in this study, the molecule interacts with a LG beam only via the excitation pathway. This scattering process depends on the inherent coupling of the excited molecule to the vacuum field to spontaneously scatter a photon. However, the set of Maxwell's equations used to derive the incident and scattered fields determine which modes are allowed to express the modes of the incident and scattered fields. On one hand, as pointed out in Section \Rmnum{2}.A, the set of LG modes are solutions to the \emph{paraxial} wave equation, which comes from the set of Maxwell's equations assuming the paraxial approximation. On the other hand, the spontaneously emitted field must solve the full, non-paraxial set of Maxwell's equations. Therefore, the LG basis set cannot describe the modes of the vacuum field into which the scattered field is spontaneously emitted. Physically, this means there is no expectation that the transverse, radial variation of the electric field couples to the vibrational degrees of freedom of the molecule during the emission transition. 

To better appreciate the physics of the LG beam-excited spontaneous scattering process, let us explicitly examine the two matrix elements of the interaction operator present in Eq. (33), $V_{ab}$ and $V_{bc}$. To do so, we assume that the model molecule possesses a single vibration. For an incident, classical $p=0$, $\ell=1$ LG beam, these matrix elements become,
\begin{subequations}
\begin{align}
V_{ab}=\langle a|H_{int}^{LG}|b\rangle=\sum_{m,\alpha}\sum_{\sigma=0,\pm1}\nonumber\\
\times\left[\mathcal{A}_{10}\left(|\vec{v}_j|/|\bar{r}_{nuc}|\right)\langle\nu_a|Q|\nu_b\rangle\langle g|E_\sigma q_{m\alpha}r_{in}Y_1^\sigma(\hat{r})|e\rangle\right.\nonumber\\\left.+\mathcal{A}_{11}\langle\nu_a|\nu_b\rangle\langle g|E_\sigma q_{m\alpha}r_{in}Y_1^\sigma(\hat{r})|e\rangle\right],\\
V_{bc}=\langle b|H_{int}^{LG}|c\rangle=\sum_{m,\alpha}\sum_{\sigma=0,\pm1}\nonumber\\
\times\left[\mathcal{A}_{10}\left(|\vec{v}_j|/|\bar{r}_{nuc}|\right)\langle\nu_b|Q|\nu_c\rangle\langle e|E_\sigma q_{m\alpha}r_{in}Y_1^\sigma(\hat{r})|g\rangle\right.\nonumber\\\left.+\mathcal{A}_{11}\langle\nu_b|\nu_c\rangle\langle e|E_\sigma q_{m\alpha}r_{in}Y_1^\sigma(\hat{r})|g\rangle\right].
\end{align}
\end{subequations}
However, as we stated above, the emission of light causing the transition from $|b\rangle$ and state $|c\rangle$ occurs due to coupling between the excited molecule and the fluctuations of the vacuum electromagnetic field. Since the field spontaneously emitted by the excited molecule must solve the full set of non-paraxial Maxwell's equations, we know that the components of the lowest order contribution to the transverse mode of the vacuum field possess a radial variation unlike that of a LG beam. This means that emission process cannot embed normal coordinates into the interaction Hamiltonian describing the emission process. To force our model to reproduce the physics expected from the lowest order contribution spontaneous scattering processes, we assume the interaction Hamiltonian associated with the operator $V_{bc}$ represents the emission of a plane wave, or a $p=0$ and $\ell=0$ LG beam. Therefore, this matrix element becomes, 
\begin{align}
V_{bc}=\sum_{m,\alpha}\sum_{\sigma=0,\pm1}\mathcal{A}_{00}\langle\nu_b|\nu_c\rangle\langle e|E_\sigma q_{m\alpha}r_{in}Y_1^\sigma(\hat{r})|g\rangle.
\end{align}

Now that we have used the third order response function to derive the analytical expression of the inelastic scattering of a classical LG beam, we must explore the form of the differential scattering cross-section for specific states of our model vibronic molecule assuming plane wave emission from the excited molecule. Using the general model vibronic states from Section \Rmnum{2}.B for the scenario of negligible pure dephasing, the differential scattering cross-section becomes, 
\begin{widetext}
\begin{align}
\sigma(\omega_L,\omega_S)=\frac{8\pi\omega_L\omega_S^3}{9\hbar^2c^4}\sum_{m,\alpha}\sum_{\sigma=0,\pm1}|M_{eg}^\sigma|^4\nonumber\\\times\sum_{\nu_b}\left|\frac{\mathcal{A}_{10}\mathcal{A}_{00}\left(\frac{|\vec{v}_j|}{|\bar{r}_{nuc}|}\right)\langle\nu_a|Q|\nu_b\rangle\langle\nu_b|\nu_c\rangle+\mathcal{A}_{11}\mathcal{A}_{00}\langle\nu_a|\nu_b\rangle\langle\nu_b|\nu_c\rangle}{\omega_{ba}-\omega_L+\imath\Gamma}\right|^2\delta(\omega_{ac}+\omega_S-\omega_L).
\end{align}
where $M_{eg}^\sigma=\langle e|E_\sigma q_{m\alpha}r_{in}Y_1^\sigma(\hat{r})|g\rangle$ is the adiabatically separated electronic transition dipole moment that participates in both the excitation and emission pathways of the scattering process and we now sum over vibrational intermediate states, $|\nu_b\rangle$ . 

Eq. (36) highlights an important facet of the inelastic scattering process resonantly excited by the incident, monochromatic and classical $p=0$, $\ell=1$ LG beam. A new term appears in the differential scattering cross-section due to the direct coupling between the radial variation of the beam and the vibrational degrees of freedom of the molecule. The magnitude of this term depends sensitively on the quantity $\mathcal{A}_{10}(|\vec{v}_j|/|\bar{r}_{nuc})\propto(v_j/w_0)=D$ where $|\vec{v}_j|=v_j$. Using the parameter $D$, the \emph{transverse coupling constant}, we approximate Eq. (36) as
\begin{align}
\sigma(\omega_L,\omega_S)=\frac{8\pi\omega_L\omega_S^3}{9\hbar^2c^4}\sum_{m,\alpha}\sum_{\sigma=0,\pm1}|M_{eg}^\sigma|^4|\mathcal{A}_{00}|^4\nonumber\\\times\sum_{\nu_b}\left|\frac{\sqrt{4/3}D\langle\nu_a|Q|\nu_b\rangle\langle\nu_b|\nu_c\rangle+\sqrt{2/3}\langle\nu_a|\nu_b\rangle\langle\nu_b|\nu_c\rangle}{\omega_{ba}-\omega_L+\imath\Gamma}\right|^2\delta(\omega_{ac}+\omega_S-\omega_L).
\end{align} 
where we have made the approximation that the longitudinal coupling between an incident field and the molecule does not differ in the cases of a plane wave versus that of a LG beam. Physically, this approximation is reasonable since the longitudinal coupling is dominated by the wavelength of the incident field relative to the position of the interacting electron and depends weakly on the beam's transverse profile.

We now use the vibronic overlaps specified in Eqs. (16a)-(16d) from Section \Rmnum{2}.B to simulate the effects of an incident, classical LG beam on the resonantly-enhanced Raman scattering process. For these states, the differential scattering cross-section becomes, 
\begin{align}
\sigma(\omega_L,\omega_S)=\frac{8\pi\omega_L\omega_S^3}{9\hbar^2c^4}\sum_{m,\alpha}\sum_{\sigma=0,\pm1}|M_{eg}^\sigma|^4|\mathcal{A}_{00}|^4\nonumber\\\times\left|\frac{\sqrt{4/3}D|\langle0|Q|0\rangle|e^{\imath\theta_1}|\langle0|1\rangle|e^{\imath\theta_2}+\sqrt{2/3}|\langle0|0\rangle|e^{\imath\theta_3}|\langle0|1\rangle|e^{\imath\theta_4}}{-\Delta+\imath\Gamma}\right.\nonumber\\\left.+\frac{\sqrt{4/3}D|\langle0|Q|1\rangle|e^{\imath\theta_5}|\langle1|1\rangle|e^{\imath\theta_6}+\sqrt{2/3}|\langle0|1\rangle|e^{\imath\theta_7}|\langle1|1\rangle|e^{\imath\theta_8}}{-(\Delta-\omega_\nu)+\imath\Gamma}\right|^2\delta(\omega_\nu+\omega_S-\omega_L).
\end{align}
\end{widetext}
where $\Delta=\omega_L-\omega_{eg}$ and we have assumed that the population of both participating excited vibronic states relax at the same rate, $\Gamma$. In this case, $\omega_{eg}$ corresponds to the frequency of the 0-0 transition from the ground to excited electronic state. Since each of the vibrational overlaps involved in the scattering process are complex numbers, Eq. (38) explicitly shows the real magnitude of this number and its associated phase. This form of the differential scattering cross-section allows us to examine how the relative phase of the two competing emission processes affects its spectrum as a function of the detuning of the incident laser frequency, $\omega_L$, from each vibronic resonance condition.   

The form of Eq. (38) is important since it possesses a change in the probability of making a scattering transition due to the functional form of the LG interaction Hamiltonian. Direct comparison of Eq. (38) to the Franck-Condon or Albrecht A-term of the resonantly enhanced inelastic light scattering differential cross-section of a molecule excited by an incident plane wave physically highlights this change.\cite{albrecht,Spiro} In the plane wave case, the A-term of this scattering cross-section is solely dependent of the Franck-Condon vibrational overlaps in the excitation and emission processes. Either type of vibrational symmetry, totally or non-totally symmetric, can cause non-zero Franck-Condon overlaps, albeit through different physical mechanisms. Excited electronic states are often displaced along totally symmetric vibrations due to bond lengthening while Jahn-Teller distorted electronic states may be displaced along specific non-totally symmetric vibrations of the molecule of interest. This similarity allows one to apply the same physical model of molecular absorption to both totally and some non-totally symmetric vibrations, as explained in Section \Rmnum{2}.B.

\begin{figure}
\centerline{\includegraphics[width=8.5 cm]{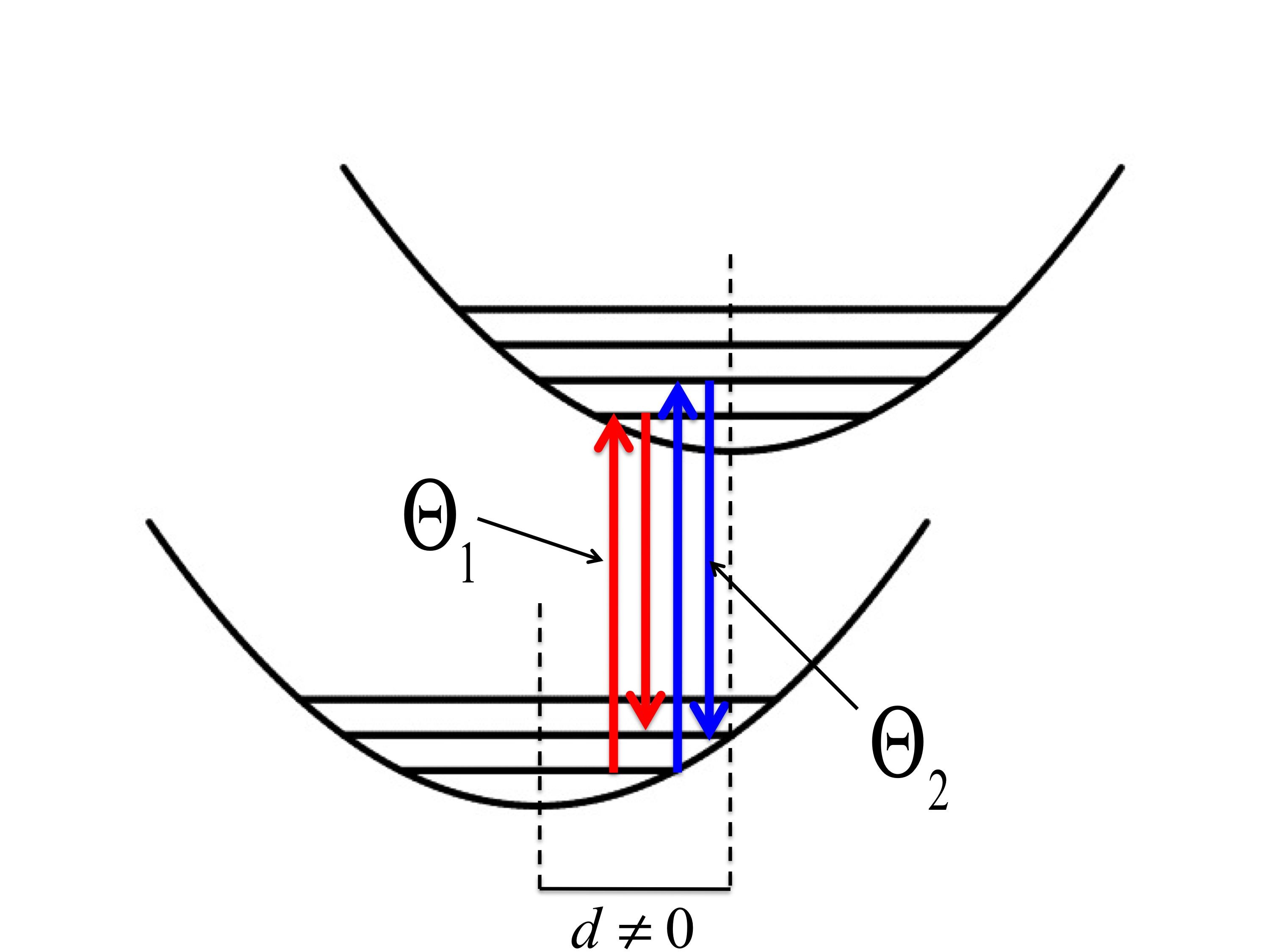}}
\caption{\label{FIG. 3}(color online) Quantum mechanical pathways that participate in resonant inelastic light scattering by model vibronic molecule explored in Section \Rmnum{2}.B. Each pathway is modulated by the coupling between incident LG beam and excited molecule, but the pathway $\Theta_2$ (right set of arrows) is enhanced relative to the intensity through $\Theta_1$ (left set of arrows) for most values of the dimensionless displacement of the excited electronic state, $\bar{d}$.}
\end{figure}

When one uses a classical LG beam to excite the electronic transition, however, we find that Eq. (38) possesses vibrational transition matrix elements that contain an embedded quantum of the normal coordinate of the model molecule due to the form of the LG-molecule interaction Hamiltonian derived in Section \Rmnum{2}.A. The appearance of the normal coordinate, $Q$, in the interaction Hamiltonian amends the transition matrix elements describing the excitation to the participating intermediate vibronic states. An embedded quantum of this model molecule's vibration in these transition matrix elements affects their value relative to the Franck-Condon overlaps. Changes to the numerical value of these matrix elements change the weight with which different quantum mechanical pathways contribute to the scattering process relative to plane wave excitation. 

\begin{figure*}[ht!]
\begin{center}
\centerline{\includegraphics[width=18.5 cm]{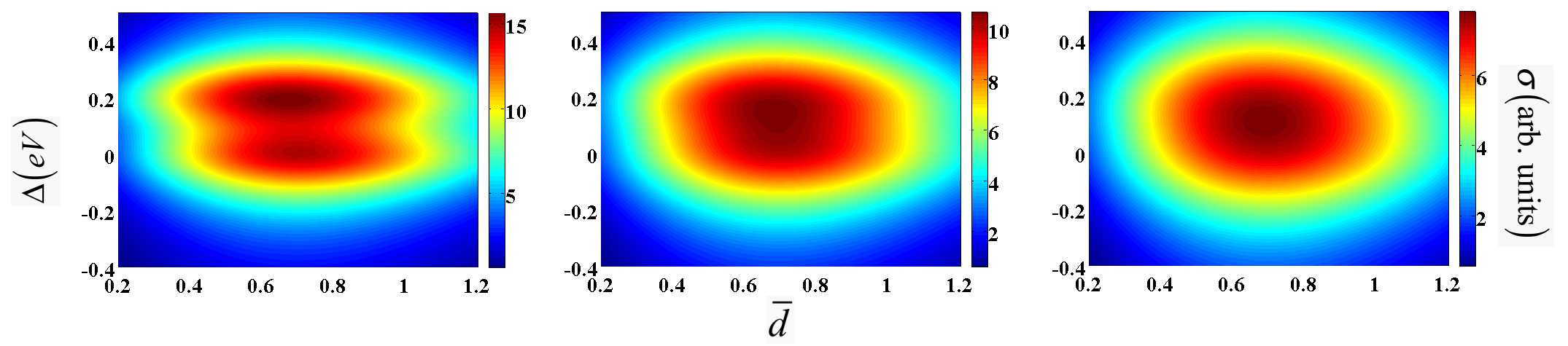}}
\caption{\label{FIG. 4}(color online) The in-phase differential resonance Raman scattering cross-section, $\sigma$ as a function of laser detuning ($\Delta$) and the dimensionless displacement of the excited electronic potential ($\bar{d}$) for a vibration of the same frequency as the C-C aromatic ring vibration of naphthalene, 1593 cm$^{-1}$, and an electronic absorption linewidth ($\Gamma$) of 0.15 eV (left), 0.2 eV (middle), and 0.25 eV (right) for the case that the coefficient $D=0.1$.}
\end{center}
\end{figure*}  

Due to this change in the participating vibrational transition matrix elements, one physical mechanism dominates changes induced by an incident LG beam in resonantly enhanced light scattering: quantum interference. Since determining the probability of making a scattering transition necessitates the coherent square over the participating quantum mechanical pathways, interference between these pathways plays an important role in resonant Raman scattering. Interference phenomena in resonance Raman scattering and Raman-like processes in molecules, molecular solids and even nano-optomechanical systems in regions of the EM spectrum ranging from microwaves to x-rays has been treated theoretically and experimentally for several decades and continues to be an important field of study within the optics community.\cite{Cardona1985,Hochstrasser,painter,Skytt_PRA_1997,Sonnich} 

For plane wave excitation, there is a finite set of intermediate vibronic states that contribute to the scattering process. The size of this set of states is determined by the difference in the structures of the ground and participating excited electronic states. As stated above, different physical mechanisms may lead to displacement of an excited electronic state along totally or non-totally symmetric vibrations. The vibronic model molecule presented in Section \Rmnum{2}.B is an example of either type of vibrational symmetry under the condition that such a displacement exists along one of the molecule's vibrations. For smaller displacements, a small set of intermediate states will contribute non-zero Franck-Condon overlaps to the scattering process. A larger displacement increases the number of intermediate states whose Franck-Condon overlap is  large enough in magnitude to make sizable contributions.  

However, at any level of displacement the $l=1$ LG beam embeds a quantum of the normal coordinate $Q$ in the first set of vibrational transition matrix elements in each of the two terms within the brackets of Eq. (38). This embedded vibrational quantum changes how vibronic states participate in the scattering process. Instead of the molecule being entirely dependent on the value of the Franck-Condon overlaps to determine the weight with which intermediate states and scattering pathways contribute, the excitation beam itself can set the value of the vibrational overlaps and fundamentally change the nature of the scattering process. Changes to the weighting of different quantum mechanical pathways to specific vibronic final states modulate the interference taking place in a scattering measurement, as seen below. 
%\begin{figure}
%\centerline{\includegraphics[width=7.5cm]{500_scattering.png}}
%\caption{\label{FIG. 2} The in-phase resonance Raman differential scattering cross-section as a function of laser detuning ($\Delta$) for a 500 cm$^{-1}$ vibration and a linewidth ($\Gamma_0$) of 3.4 THz for three values of the coefficient $D$: $D=0.1$ (blue), $D=0.01$ (red), and $D=0.001$ (green). The inset shows an enlarged view of the region near zero detuning where the magnitude of the interference contribution becomes negative.}
%\end{figure}

The interference present in the scattering process comes from two categories of sources. The first category represents properties of the molecule that the experimenter cannot change. These properties include the relative phase of the two emission vibronic transition moments and the relaxation rates of the participating excited vibronic states. The second category includes the frequency of the laser relative to the frequency of the electronic transition ($\Delta=\omega_L-\omega_{eg}$), the frequency of the vibration of interest, and the mode of the incident electromagnetic laser field. Both the relative phase of the transition matrix elements and the energies of its stationary, unperturbed core states are fixed by the atoms of the molecule. However, the researcher controls the energy of the light beam and the frequency of the vibration of interest, as well as the transverse mode of the incident beam. Using a $p=0$ , $l=1$ LG beam at specific detunings from the resonant transition frequencies and examining specific vibrational frequencies, a researcher can change the sign and magnitude of the modulation to the interference taking place in the scattering process. This point is clearer when examining the dependence of the differential scattering cross-section as a function of incident laser frequency, different excited state population relaxation rates and the dimensionless displacement of the excited electronic potential, $\bar{d}$.

The two scattering pathways we explicitly examine in Eq. (38) using our vibronic model molecule are visualized in Figure 3.
We denote the two pathways as $\Theta_1$ and $\Theta_2$. In terms of the states defined above in Section \Rmnum{2}.B, the path of $\Theta_1$ is $|g\rangle|0\rangle\rightarrow|e\rangle|0\rangle\rightarrow|g\rangle|1\rangle$ while that of $\Theta_2$ is $|g\rangle|0\rangle\rightarrow|e\rangle|1\rangle\rightarrow|g\rangle|1\rangle$. For specific values of vibrational frequency, $\omega_\nu$, and population relaxation rate, $\Gamma$, these two pathways interfere and modulate the intensity of the classical LG beam spontaneously scattered from the molecule.  

Figure 4 shows a 2D plot of the differential resonance Raman scattering cross-section in Eq. (38) for vibrational frequencies equal to the totally symmetric C-C ring vibration of naphthalene (1593 cm$^{-1}\approx0.197$ eV) as a function of the laser detuning ($\Delta$) and the dimensionless displacement of the excited electronic potential ($\bar{d}$) when the transverse coupling constant $D=0.1$ and the two emission transition moments are \emph{in-phase} with each other for three different transition linewidths: 0.15 eV (left), 0.20 eV (middle) and 0.25 eV (right). 
\begin{figure*}[ht!]
\begin{center}
\centerline{\includegraphics[width=18.5 cm]{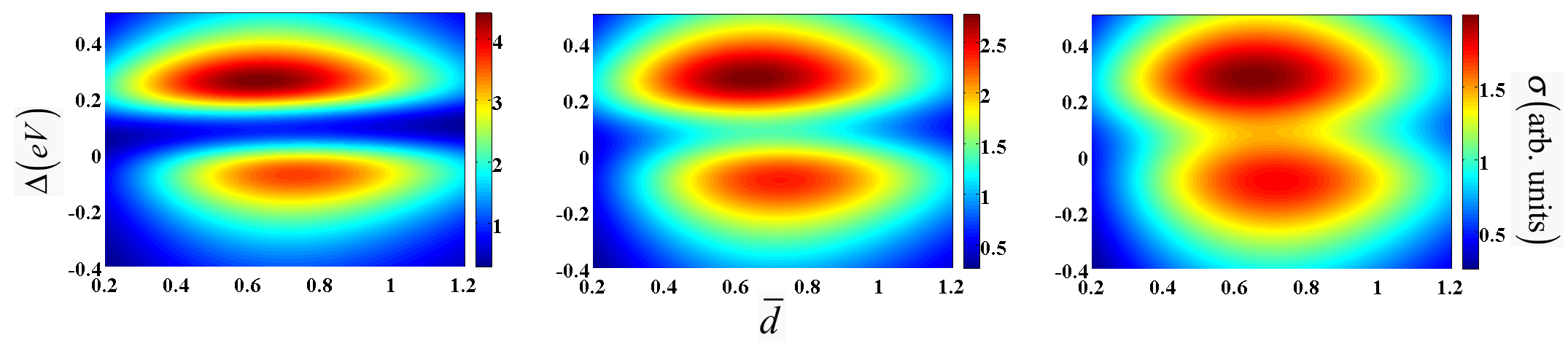}}
\caption{\label{FIG. 5} (color online) out-of-phase differential resonance Raman scattering cross-section, $\sigma$ as a function of laser detuning ($\Delta$) and the dimensionless displacement of the excited electronic potential ($\bar{d}$) for a vibration of the same frequency as the totally symmetric C-C ring vibration of naphthalene, 1593 cm$^{-1}$, and an electronic absorption linewidth ($\Gamma$) of 0.15 eV (left), 0.2 eV (middle), and 0.25 eV (right) for the case that the coefficient $D=0.1$.}
\end{center}
\end{figure*}
In contrast, Figure 5 shows the dependence of the differential resonance Raman scattering cross-section in Eq. (38) for the C-C ring vibration of naphthalene as a function of the laser detuning and displacement of the participating excited state for the case that the two emission transition moments are \emph{out-of-phase} and the transverse coupling constant $D=0.1$ for the same three excited state absorption linewidths as Figure 4. This figure shows that a detuning frequency near $\Delta=\omega_\nu/2$ leads to a minimum in the intensity scattered by this model molecule. Stroud and co-workers found a similar result when examining a fully quantum mechanical description of the interaction for an electric field with a multi-level atomic system and it represents a steady-state destructive quantum mechanical interference phenomenon between the two scattering pathways.\cite{Stroud} It is encouraging that the semi-classical method used in the treatment presented here qualitatively matches the physics uncovered in that previously published, fully quantum mechanical treatment.

\begin{figure}
\centerline{\includegraphics[width=7.5 cm]{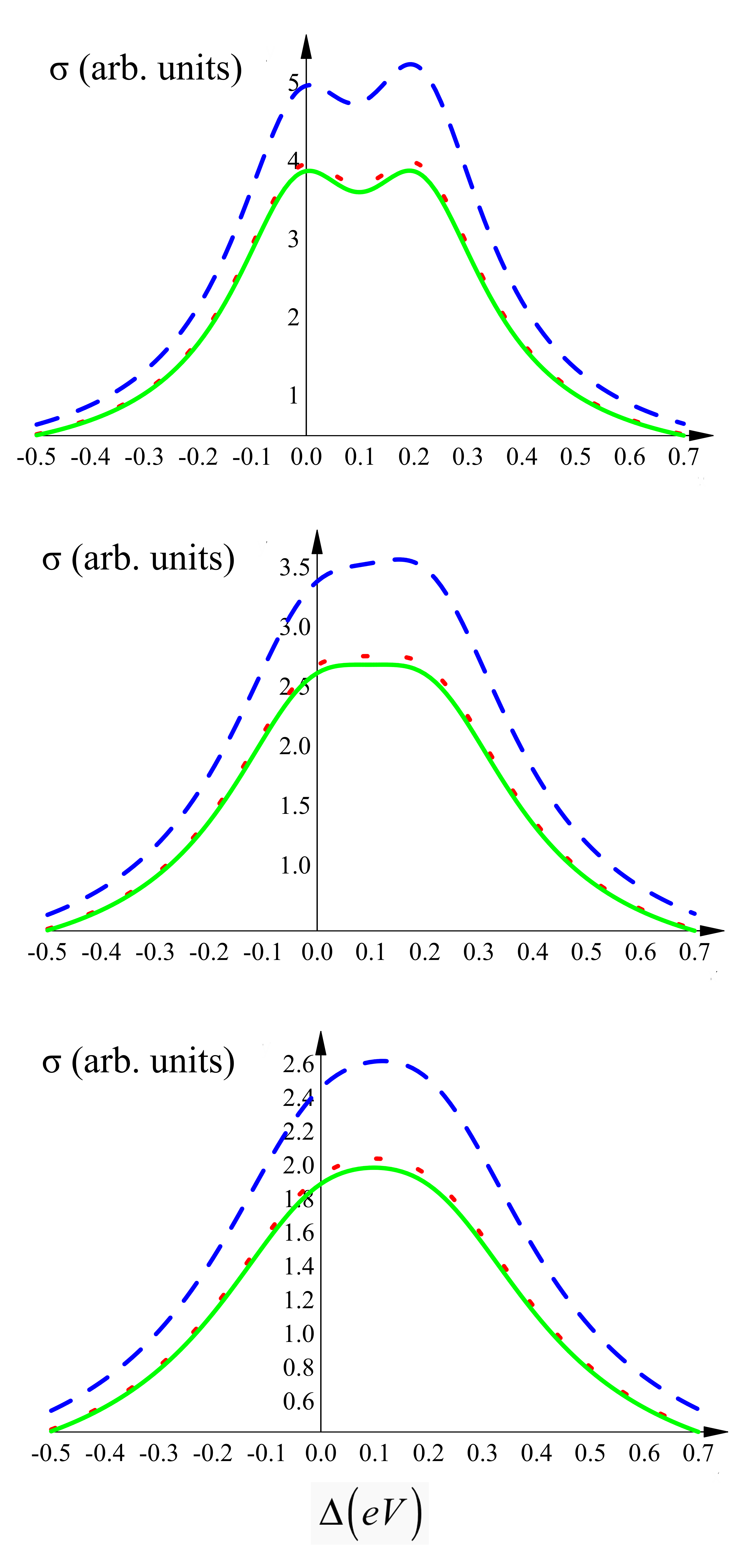}}
\caption{\label{FIG. 6}(color online) The in-phase differential resonance Raman scattering cross-section as a function of laser detuning ($\Delta$) in the case that $\bar{d}=0.7$ for three different electronic absorption linewidths $\Gamma=0.15$ eV (top), $\Gamma=0.20$ eV (middle), and $\Gamma=0.25$ eV (bottom) and three values of the transverse coupling constant $D$: $D=0.1$ (dashed), $D=0.01$ (dotted), and $D=0.001$ (solid).}
\end{figure}  

%\begin{figure}
%\centerline{\includegraphics[width=7.5cm]{250_scattering.png}}
%\caption{\label{FIG. 3} The in-phase resonance Raman differential scattering cross-section as a function of laser detuning ($\Delta$) for a 250 cm$^{-1}$ vibration and a linewidth ($\Gamma_0$) of 3.4 THz for three values of the coefficient $D$: $D=0.1$ (blue), $D=0.01$ (red), and $D=0.001$ (green).}
%\end{figure}
%\begin{figure}
%\centerline{\includegraphics[width=7.5cm]{1500_scattering_phase_out.png}}
%\caption{\label{FIG. 4}The out-of-phase differential resonance Raman scattering cross-section as a function of laser detuning ($\Delta$) for a 1500 cm$^{-1}$ vibration and a linewidth ($\Gamma_0$) of 3.4 THz for three values of the coefficient $D$: $D=0.1$ (blue), $D=0.01$ (red), and $D=0.001$ (green).}
%\end{figure}

Three aspects of the differential resonance Raman scattering cross-section stand out in Figures 4 and 5. First, five of the six spectra presented show an asymmetry in peak intensity when considering each participating scattering pathway. In these five spectra the scattering intensity through the $\Theta_2$ pathway is larger than that through $\Theta_1$. In their treatment of the steady-state interference present in resonantly-enhanced light scattering processes, Stroud and co-workers showed a symmetry with respect to the scattered intensity when the incident laser frequency matched that of the resonant transitions of a multi-level atom.\cite{Stroud} That is, in that study the predicted scattering intensity did not depend on the excitation of specific transitions. The peak scattering intensity through one quantum mechanical pathway equals that through each of the other resonant transition in a multi-level atom with plane wave excitation. However in the present case, the coupling of the classical LG beam to the vibrational degree of freedom of a model molecule destroys the symmetry in the value of the matrix elements determining the intensity with which each pathway participates in the scattering process. This coupling favors the $\Theta_2$ pathway, as witnessed by its enhancement relative to the $\Theta_1$ pathway in the LG-molecule scattering spectra shown in Figures 4 and 5. Only the in-phase spectra with an absorption linewidth of $\Gamma=0.25$ eV is almost totally symmetric with respect to the tuning of the incident laser frequency to each resonant vibronic transition. 

Second, the position and shape of the vibronic peaks present in the LG-amended differential resonance Raman scattering cross-section depend on both the lifetime of the participating excited state and the relative phase of the two participating vibronic emission transition moments. For the in-phase case shown in Figure 4, the symmetry of the differential cross-section as a function of both the laser detuning and dimensionless displacement parameter depend sensitively on the absorption linewidth. The spectrum for the $\Gamma=0.25$ eV case is the most symmetric while that of $\Gamma=0.15$ eV case is least symmetric. In the case of a $\Gamma=0.25$ eV linewidth, it is difficult to distinguish the two distinct vibronic scattering pathways, $\Theta_1$ and $\Theta_2$, within the single peak feature present in the spectrum. In the spectra for all three linewidths examined above, the peak corresponding to the $\Theta_2$ pathway has shifted down in energy from a theoretically anticipated position peak of $\Delta=\omega_\nu$ eV to a lower value while that corresponding to the $\Theta_1$ pathway has shifted up in energy from an anticipated position of $\Delta=0$ eV to a larger value. In the case of an absorption linewidth of $\Gamma=0.25$ eV, this effect produces a single, mostly symmetric differential scattering lineshape with respect to both the laser detuning frequency and displacement of the participating excited vibronic states. In the case of each of the two more narrow linewidths (longer-lived excited states) there is an asymmetry along the dimensionless displacement, $\bar{d}$, with the peak associated with the $\Theta_2$ pathway maximizing at smaller values of $\bar{d}$ than that of the $\Theta_1$ pathway. This asymmetry is more noticeable in the case of $\Gamma=0.15$ eV than that of $\Gamma=0.2$ eV, as seen below.

\begin{figure}
\centerline{\includegraphics[width=7.5cm]{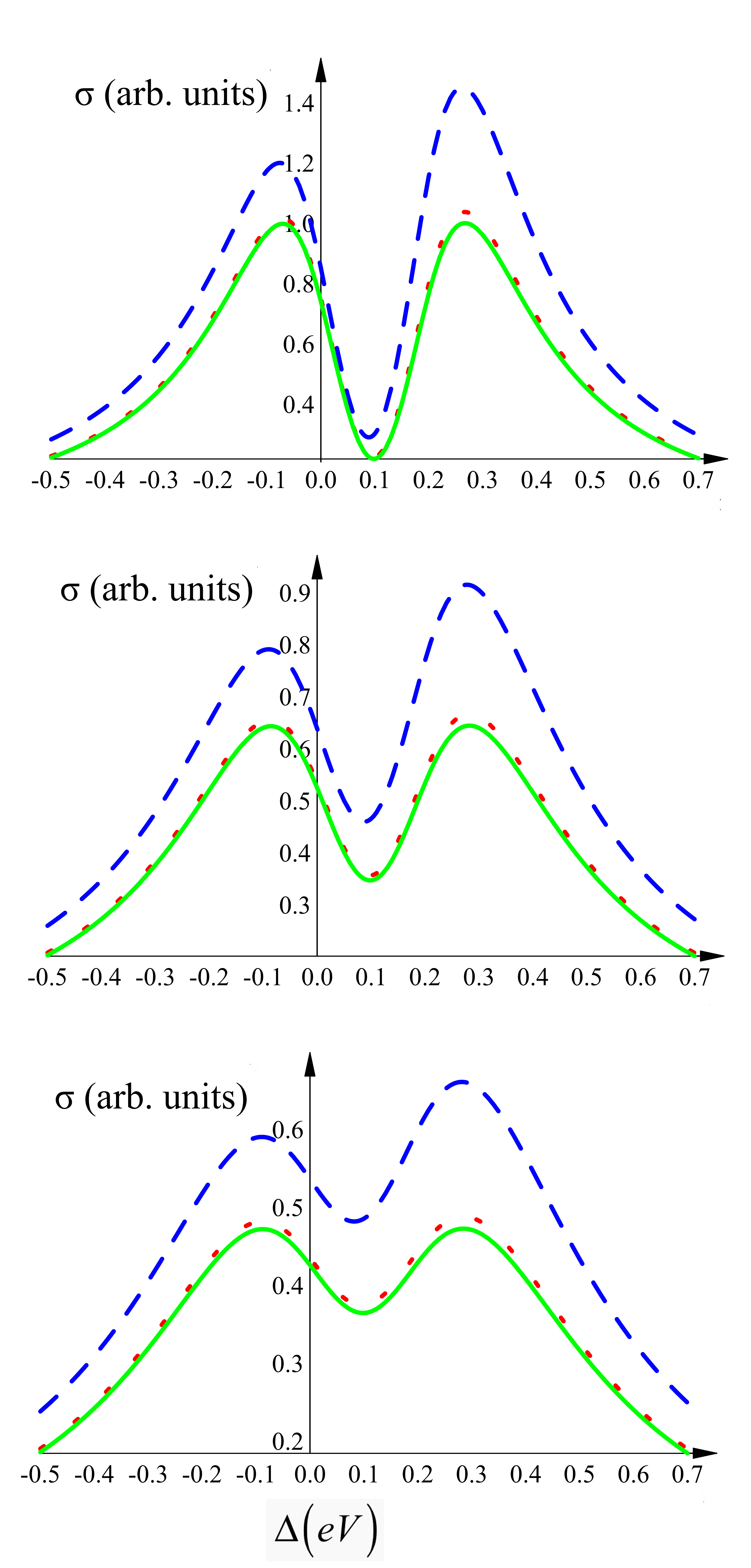}}
\caption{\label{FIG. 7}(color online) The out-of-phase differential resonance Raman scattering cross-section as a function of laser detuning ($\Delta$) in the case that $\bar{d}=0.7$ for three different electronic absorption linewidths $\Gamma=0.15$ eV (top, $\Gamma=0.20$ eV (middle), and $\Gamma=0.25$ eV (bottom) and three values of the transverse coupling constant $D$: $D=0.1$ (dashed), $D=0.01$ (dotted), and $D=0.001$ (solid).}
\end{figure}

In contrast to the shifts in Figure 4, Figure 5 shows that when the two emission vibronic transitions are out-of-phase, each vibronic resonance peak in the differential scattering cross-section shifts in the opposite direction along the detuning frequency axis. The peak for the $\Theta_2$ pathway shifts to larger, positive laser detunings while the peak for the $\Theta_1$ pathway shifts to larger, negative detunings. Also in contrast to Figure 4, Figure 5 shows that this shift is smallest for the case when the absorption linewidth is $\Gamma=0.25$ eV. In that case, the peak associated with the $\Theta_2$ pathway shifts to a laser detuning frequency near $\Delta=0.3$ eV, significantly larger than the position anticipated from the Lorentzian form of second term in Eq. (38) while the peak associated with the $\Theta_1$ pathway shifts to a detuning more negative than $\Delta=-0.1$ eV. While these shifts are in the same direction in the case of the more narrow lineshapes ($\Gamma=0.2$ eV and $\Gamma=0.15$ eV), these shifts are larger in magnitude and depend explicitly on the LG-molecule coupling. The case of each phase relationship between the respective vibronic emission transition moments do show a similar trend when examining the dependence of the lineshape of differential scattering cross-section of the dimensionless displacement. The most narrow absorption linewidth ($\Gamma=0.15$ eV) shows the most asymmetry along this direction while the widest linewidth ($\Gamma=0.25$ eV) shows the least asymmetry when these moments are either in or out-of-phase with one another.   

Third, Figures 4 and 5 also show similar trends in the intensity of the peaks present in the differential scattering cross-section spectra. Whether the emission transitions are in or out-of-phase, the most intense spectra correspond to the physical situation in which the linewidth is the smallest. This conclusion is easily understood when examining Eq. (38). When the vibronic resonance condition is met (either $\Delta=0$ eV or $\Delta=\omega_\nu$ eV), only the value of the absorption linewidth is left in the denominator of the first two terms in Eq. (38). As this value becomes smaller, one expects the intensity of scattered light to increase. The spectra in Figures 4 and 5 confirm this expectation.

\begin{figure}
\centerline{\includegraphics[width=9 cm]{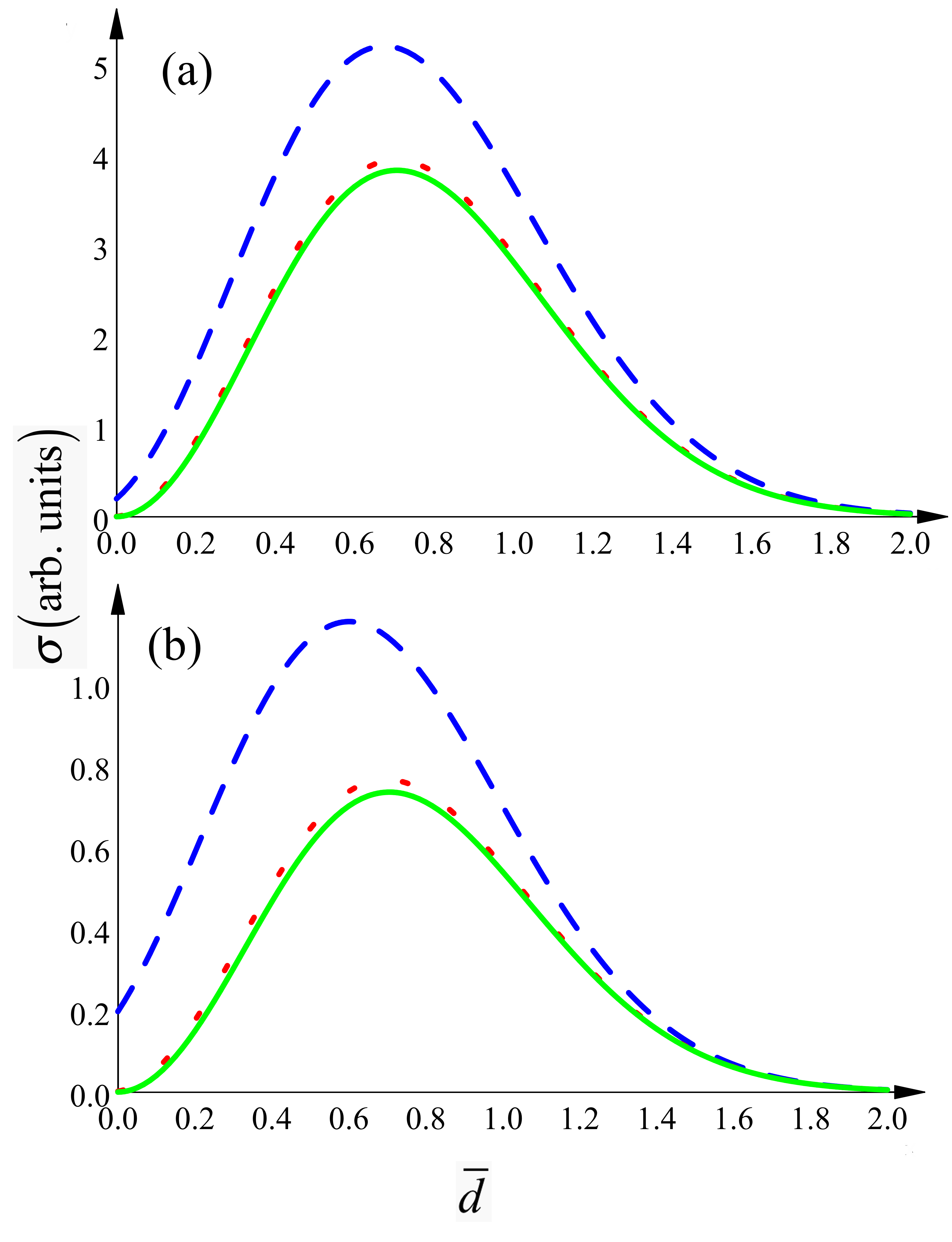}}
\caption{\label{FIG. 8}(color online) The effect of the transverse coupling of the incident LG beam and vibronic molecule on the dependence of the differential resonant inelastic scattering cross-section of $\bar{d}$ for the case that $\Delta=\omega_\nu=0.197$ eV, the transition moments of the two contributing emission process are in-phase (a) and out-of-phase (b) and three values of the transverse coupling constant $D$: $D=0.1$ (dashed), $D=0.01$ (dotted), and $D=0.001$ (solid).}
\end{figure}

The difference in symmetry of the vibronic peaks of the differential scattering cross-section spectra in Figures 4 and 5 allows one to determine the capability of the the incident LG to modulate quantum mechanical interference present in the scattering process. When the linewidth is significantly larger than the vibrational frequency of interest and the two emission transition moments are in-phase, the differential resonance Raman scattering spectrum is almost totally symmetric with respect to both the frequency of the incident laser field and the displacement of the excited electronic state. This means that the interference between the different scattering pathways not only shifts the position of the peak scattering intensity spectrally (with respect to $\Delta$) but also structurally (with respect to displacement of the excited state). This conclusion is supported by the fact that the differential scattering cross-section is asymmetric along the structural parameter $\bar{d}$ for the two more narrow linewidths examined in Figure 4. Physically, this structural symmetry corresponds to an interferometric mechanism whereby the two scattering pathways equally share resonant intensity via the absorption feature broader than the frequency of the pertinent vibration and intensity symmetrically interferes as a function of the displacement of the molecule's excited electronic state. 

However, the observed symmetry in this spectrum of Figure 4 does \emph{not} change when one varies the strength of the coupling between the incident LG beam and the vibration of the model molecule. Figure 6 shows a slice of the 2D in-phase differential cross-section when $\bar{d}=0.7$, near the the peak of the scattering intensity through the $\Theta_2$ pathway, for three different values of the transverse coupling constant, $D$. This figure shows that while the overall scattered intensity increases as $D$ increases, the lineshape of the spectrum does not. The spectrum stays symmetric with respect to the detuning of the incident field from the participating resonant vibronic transitions and must be a natural consequence of the larger than 1 ratio of the linewidth of the resonant absorption feature to the vibrational frequency of interest. In contrast, the spectra in cases of absorption linewidths of $\Gamma=0.15$ eV and  $\Gamma=0.20$ eV both become more asymmetric with respect to $\Delta$ as $D$ increases. However, the positions of the peaks associated with each respective scattering pathway show little to no variation as a function of the magnitude of the coupling constant.  

%\begin{figure}
%\centerline{\includegraphics[width=7.5cm]{500_scattering_phase_out.png}}
%\caption{\label{FIG. 5} The out of phase differential resonance Raman scattering cross-sectionas a function of laser detuning ($\Delta$) for a 500 cm$^{-1}$ vibration and a linewidth ($\Gamma_0$) of 3.4 THz for three values of the coefficient $D$: $D=0.1$ (blue), $D=0.01$ (red), and $D=0.001$ (green).}
%\end{figure} 

The situation differs significantly when the two emission transition moments are out-of-phase. Figure 7 shows a slice of the 2D out-of-phase differential cross-section when $\bar{d}=0.7$ and for the same three values of the transverse coupling constant, $D$, as in Figure 6. In that case, all of the differential scattering cross-section spectra possess spectral asymmetry. However, unlike Figure 6, the position of both the peak corresponding to the $\Theta_1$ pathway and the intensity minimum shift to more negative detuning frequencies as a function of the transverse coupling constant for two of the spectra presented in Figure 7. For absorption linewidth of $\Gamma=0.15$ eV and  $\Gamma=0.20$ eV these shifts are visible as the value of $D$ increases from 0.001 to 0.1. A shift may also be present for the case of of $\Gamma=0.25$ eV, but it is not as clear in the simulated spectra as in the case of smaller linewidths. These spectra clearly show that the strength of the coupling of the transverse variation of the incident electromagnetic mode and the vibrational degree of freedom of the molecule modulates the quantum mechanical interference taking place in the scattering process. 

One also sees that as $D$ approaches smaller values, the detuning spectra in Figures 6 and 7 become increasing symmetric with respect to the scattered intensity through each participating pathway. Physically, reducing the transverse coupling constant to $D=0.001$ reproduces what is expected in the case of plane wave excitation of the scattering process and we recover the symmetry first predicted by Stroud and co-workers explained above when $\Gamma=0.15$ eV. Again, this recovery of results predicted from a fully quantum mechanical treatment gives confidence to the method used in this study and the results we derive from it. 

Perhaps as interesting as the shift of the $\Theta_1$ peak of the $\Delta$-dependent differential scattering cross-section spectra in Figure 7, there is also a shift in the peak of the differential scattering cross-section as a function of the dimensionless displacement, $\bar{d}$. This shift is observed in Figure 8 which shows the dependence of $\sigma$ as a function of $\bar{d}$ for the an absorption linewidth of $\Gamma=0.15$ eV for both the cases that the emission transition moments are in and out-of-phase for the same three values of the transverse coupling constant when $\Delta=\omega_\nu=0.197$ eV, simulating the C-C ring breathing vibration of naphthalene. One observes that the displacement of the excited state at which the intensity of the scattering light peaks shifts to smaller values. However, this shift is not caused by a change in the structure of the molecule, but rather an actively tunable coupling between the spatial structure of the incident light mode and the molecule's quantized vibrational degrees of freedom. Such a shift would represent the first method available that allows researchers to use a light field to match the spectral peak of resonant processes to structural parameters of molecular samples to maximize measured signal.

%\begin{figure}
%\centerline{\includegraphics[width=8.5 cm]{phase_compare.png}}
%\caption{\label{FIG. 7}(color online) (top panel) The excitation-emission relative phase dependence of the differential resonance Raman scattering cross-section of a model vibronic molecule possessing whose dimensionless displacement is $\bar{d}=0.7$, absorption linewidth $\Gamma=0.25$ eV due vibrational frequency of 1593 cm$^{-1}$ due to an incident classical $p=0$, $\ell=1$ LG beam for three different values of the parameter $D$ from Eq. (37): $D=0.1$ (blue), $D=0.01$ (green), and $D=0.001$ (red).(bottom panel)Close-up of the region of relative phase at which the scattering intensity minimizes showing that this intensity does not reach zero in the case of strong coupling between the radial variation of the incident beam and the molecule's vibration.}
%\end{figure}

%\begin{figure}
%\centerline{\includegraphics[width=7.5cm]{250_scattering_phase_out.png}}
%\caption{\label{FIG. 6} Differential resonance Raman scattering cross-sectionas a function of laser detuning ($\Delta$) for a 250 cm$^{-1}$ vibration and a linewidth ($\Gamma_0$) of 3.4 THz for three values of the coefficient $D$:$D=0.1$ (blue), $D=0.01$ (red), and $D=0.001$ (green).}
%\end{figure}
The above analysis leads to two important conclusions concerning the feasibility of coherent control of molecular dynamics using incident LG beams to excite resonant, inelastic light scattering. First, the linewidth of the resonant feature should approximately match the molecular vibrational frequency of interest, but not exceed it. Second, the magnitude of the vibration, $v_j$, must approach at least 1/10th the size of the incident beam waist. Satisfaction of these two conditions leads to modulation of the quantum mechanical interference taking place in the scattering process, both spectrally and structurally. However, satisfaction of these conditions implies that for quantized molecular vibrations one needs a beam waist on the order of 10's of nm, at most. Only resonant scattering processes in the soft x-ray regime can fulfill both of these requirements.   

Like Raman processes in other regions of the EM spectrum, resonant inelastic x-ray scattering (RIXS) probes the symmetry and structure of molecular states as a function of their immediate environment with atomic sensitivity.\cite{XrayrR}. Quite surprisingly, despite the involvement of core electronic states that are highly localized to atoms rather than shared in molecular bonds, vibrations of molecules can modulate absorption of x-rays. When the lifetime associated with the core electron-hole pair is shorter than the period of the vibrations of interest, the vibronic overtones overlap in energy and lead to interference in the RIXS process. This effect is known as \emph{lifetime-vibrational interference}.\cite{Agren_JCP_1985,Skytt_PRA_1997,XrayrR}

A classic example of the effects of lifetime-vibrational interference is seen in the 1s K-edge absorption of the O atom in CO. By comparing x-ray absorption spectra (XAS) excited in a range of photon energies between 533.4 eV and 535.3 eV to numerical simulations, Skytt \emph{et al.} show that the several overtones of the C-O stretching frequency at 1431.1 cm$^{-1}$ overlap within the linewidth of the atomic absorption feature.\cite{Skytt_PRA_1997} Using this physical insight, they then showed that including interference between the quantum mechanical pathways into calculations of the RIXS process correctly predicted the measured spectra, while neglecting the interference terms produced simulations that poorly matched experiments. 

In contrast to 1s O excitation, Skytt \emph{et al.} found that the presence or absence of lifetime-vibrational interference effects in 1s C RIXS excitation near 287 eV depended much more sensitively on the frequency of the incident beam. When the frequency is tuned to the 0-0 vibronic core transition at 287.40 eV, no interference is observed. However, when the frequency is tuned to the 0-2 vibronic transition interference terms must be added in simulations to account for the structure of the observed RIXS spectrum. 

While lifetime-vibrational interference effects have been established in the case of small molecules like CO and N$_2$, little work explores these effects in the context of larger molecular systems.\cite{Glans_1996,Skytt_PRA_1997} However, vibrational structure has been observed in the near-edge x-ray absorption fine structure (NEXAFS) spectra of large molecules, including naphthalene.\cite{minkov:5733} Given the measured and calculated linewidths of the core electronic transitions and the frequencies of the vibrations in aromatic molecules like benzene and naphthalene as well as observation of vibrational structure in NEXAFS spectra, it is likely that quantum mechanical interference plays a significant role in determining the spectra observed in RIXS experiments of these systems. 

Based on the results of this paper and the properties of its core electron x-ray absorption, we propose an avenue to electromagnetically control the behavior observed in the RIXS of large, poly-atomic aromatic molecules. In addition to the theoretical results presented in this study, several groups have confirmed production of the x-ray vortex beams carrying well-defined angular momentum and radial variation in direct comparison to similar beams in other spectral regions.\cite{x-rayvortex3,x-rayvortex2,x-rayvortex} Also, several groups of researchers around the world have proposed and tested methods for focusing x-rays from modern sources to beam sizes tens to hundreds of nm's across.\cite{x-raynano,x-raynano2,x-raynano3} While many of these techniques focus on hard x-rays whose energy exceeds that of core electronic transitions, there is no a priori reason why similar techniques cannot be applied in the region of 200-1000 eV. 

With beam waists that small, one can imagine producing a x-ray LG beam whose radial variation couples to the internal degrees of freedom of large poly-atomic molecules and amends the resonant transitions to intermediate states in a scattering measurement. Using similar techniques, one would match the waist of a resonant soft x-ray beam to the amplitude of different molecular vibrations to directly couple these nuclear motion to the radial variation of the incident field. In doing so, experimentalists could change chemical, thermal and other environmental parameters of solid-state samples using this molecule to determine how this coupling affects reactivity, dynamics and behavior of molecular materials increasingly important to information technology, biophysics, energy science and material science. 

\section{Conclusions}
This paper develops the resonantly-enhanced inelastic scattering of a classical LG beam from the vibrations of a model vibronic molecule. The coupling between the transverse radial variation of the LG beam and vibrations of the molecule embeds units of vibrational quanta in the vibronic transition matrix elements, affecting their value relative to the case of plane wave excitation. The ability to directly affect the value of vibrational transition matrix elements corresponds to an electromagnetically mediated vibronic coupling mechanism similar to the coupling of the electronic and vibrational degrees of freedom of many molecules. In the context of resonant inelastic scattering, this electromagnetically mediated vibronic coupling changes the probability of making scattering transitions as a function of several parameters of this material system. Specifically, this coupling changes the weighting with which different quantum mechanical pathways contribute to the scattering process.

For the case of the model vibronic molecular system whose vibrational frequency of interest matches that of the C-C ring vibration of naphthalene, an incident, classical $p=0$ , $l=1$ LG beam enhances each of two participating quantum pathways leading to the final vibronic state $|e\rangle|\nu\rangle=|0\rangle|1\rangle$, but this enhancement differs for each pathway. The asymmetric enhancement modulates the differential resonance Raman scattering cross-section via both incoherent and coherent mechanisms. The modulation is dependent on the laser detuning from the resonant vibronic excitation transitions, the linewidth of the resonant spectral feature, the relative phase of the emission vibronic transition dipole moments and magnitude of the transverse coupling constant $D$, relating the amplitude of the vibration to the waist of the incident light beam.

For large values of $D$, the LG beam increases the total scattering intensity and can create an intensity asymmetry in scattering spectra studied as a function of the tuning of the incident light frequency to each participating vibronic resonance. In the case that the emission dipole moments from each respective intermediate state are in phase, this asymmetry is present for absorption linewidths of $\Gamma=0.15$ eV and $\Gamma=0.20$ eV, while the spectrum associated with a linewidth of $\Gamma=0.25$ eV is symmetric with respect to both frequency of the incident light field and displacement of the participating excited electronic state. 

When the emission dipole moments are out of phase, an intensity asymmetry appears in the case of all three of absorption linewidths examined. In the case of the two smaller linewidths ($\Gamma=0.15$ eV and $\Gamma=0.20$ eV) this intensity asymmetry appears in conjunction with LG-induced spectral shifts to the peaks of the differential scattering cross-section intensity to larger negative frequency detunings, away from the lower energy vibronic resonant transition. This shift disappears as the transverse coupling constant decreases in value, matching the qualitative features of these spectra shown in previous, fully quantum mechanical studies. 

This investigation also shows that for the case of an absorption linewidth slightly smaller than the vibrational frequency of interest, the transverse coupling between the incident field and molecule causes a shift in the value of the displacement of the participating electronic excited state at which the scattered intensity maximizes. The ability to tune this parameter of the light-matter interaction may provide researchers new information about potential energy surfaces along which electrons move in photo-excited processes and the ability to maximize measured signals through tailored coupling between the incident field and molecular vibrations.  

Simulations show an optimal value of the transverse coupling constant, $D$, near 0.1. Physically, this means that the effects of the coupling of the transverse radial variation of an incident LG beam to a molecule's vibrations are most prominent when the waist of the incident beam is 10 times the size as the amplitude of vibrations of interest. Since these amplitudes rarely exceed 1 nm, we propose that the effects of transverse coupling may be most applicable in resonant inelastic x-ray scattering (RIXS) measurements using core electron transitions for resonant enhancement. Using RIXS, we propose to test the results of this paper using 1s carbon transitions in naphthalene and other aromatic molecular samples. Based on lifetime-vibrational interference observed in small molecules including CO we believe that LG-mediated transverse coupling to vibrational degrees of freedom could provide important and novel information about molecular materials increasingly important to fields ranging from information science to biophysics to energy technologies.  

\begin{acknowledgments}
The author thanks R.J. Sension, P.R. Berman,  D.G. Steel and R.M. Freeling for useful discussions. This work was supported by the Defense Threat Reduction Agency-Joint Science and Technology Office for Chemical and Biological Defense (grant HDTRA1-09-1-0005). 
\end{acknowledgments}

%\bibliography{LG_Raman_bib.bib}
%\bibliographystyle{osajnl.bst}

%%Do not include separate BibTeX files; if BibTeX is used,
%% paste the output (contents of .bbl file) here.
}}}}}}}}}}}}}}}}}}}}}

\end{document}